\documentclass[letterpaper,twocolumn,10pt]{article}
\usepackage{usenix2019_v3}

\usepackage{tikz}
\usepackage{amsmath}

\usepackage[available, functional]{usenixbadges}
\usepackage{filecontents}
\usepackage{booktabs}
\usepackage{microtype}
\usepackage{graphicx}
\usepackage{subcaption}
\usepackage{booktabs} % for professional tables

\usepackage{amsmath}
\usepackage{multirow}
\usepackage{algorithm}
\usepackage[noend]{algpseudocode}
\usepackage{enumitem}
\usepackage{tikz}
\usepackage{amsfonts}
\usepackage[small,compact]{titlesec}
\usepackage{booktabs} % For better table lines
\usepackage{pifont}    % For checkmarks and crosses
\usepackage{xcolor}

\newcommand{\cmark}{\ding{51}}% Check mark
\newcommand{\xmark}{\ding{55}}% Cross mark

\newcommand{\edit}[1]{#1}

\makeatletter
\renewcommand\paragraph{\@startsection{paragraph}{4}{\z@}%
  {0pt} % No vertical space before
  {0pt} % No vertical space after
  {\normalfont\normalsize\bfseries}} % Keep the existing style
\makeatother

\let\oldparagraph\paragraph
\renewcommand{\paragraph}[1]{\oldparagraph{#1}\hspace{.25em}}

\begin{document}
%-------------------------------------------------------------------------------

%don't want date printed
\date{}

% \author{Jason Mohoney, Devesh Sarda, Mengze Tang, Shihabur Rahman Chowdhury*, Anil Pacaci*, Ihab F. Ilyas\textdagger, Theodoros Rekatsinas*, Shivaram Venkataraman}

% \def \authors{Jason Mohoney, Devesh Sarda, Mengze Tang, Shihabur Rahman Chowdhury*, Anil Pacaci*, Ihab F. Ilyas\textdagger, Theodoros Rekatsinas*, Shivaram Venkataraman}
% \affiliation{
%         \institution{University of Wisconsin-Madison}
%         \institution{\textdagger University of Waterloo}
%         \institution{*Apple}
%     \country{}
% }


\title{\Large \bf Quake: Adaptive Indexing for Vector Search}

% \author{Paper \#870}
%for single author (just remove % characters)
% \author{
% {\rm Your N.\ Here}\\
% Your Institution
% \and
% {\rm Second Name}\\
% Second Institution
% copy the following lines to add more authors
% \and
% {\rm Name}\\
%Name Institution
% } % end author

% \author{{Jason Mohoney} \\ University of Wisconsin-Madison \and {Devesh Sarda} \\ University of Wisconsin-Madison \and {Mengze Tang} \\ University of Wisconsin-Madison \and {Shihabur Rahman Chowdhury} \\ Apple \and{ Anil Pacaci} \\ Apple \and {Ihab F. Ilyas} \\ University of Waterloo \and {Theodoros Rekatsinas} \\ Apple \and {Shivaram Venkatarama} \\ University of Wisconsin-Madison }

\author{
  {\rm Jason Mohoney} \\ University of Wisconsin-Madison
  \and
  {\rm Devesh Sarda} \\ University of Wisconsin-Madison
  \and
  {\rm Mengze Tang} \\ University of Wisconsin-Madison
  \and
  {\rm Shihabur Rahman Chowdhury} \\ Apple
  \and
  {\rm Anil Pacaci} \\ Apple
  \and
  {\rm Ihab F. Ilyas} \\ University of Waterloo
  \and
  {\rm Theodoros Rekatsinas} \\ Apple
  \and
  {\rm Shivaram Venkataraman} \\ University of Wisconsin-Madison
}

\maketitle

\pagestyle{empty}

%-------------------------------------------------------------------------------
\begin{abstract}

Vector search, the task of finding the k-nearest neighbors of a query vector against a database of high-dimensional vectors, underpins many machine learning applications, including retrieval-augmented
generation, recommendation systems, and information retrieval. However, existing approximate nearest neighbor (ANN) methods perform poorly under dynamic and skewed workloads where data distributions evolve. We introduce Quake, an adaptive indexing system that maintains low latency and high recall in such environments. Quake employs a multi-level partitioning scheme that adjusts to updates and changing access patterns, guided by a cost model that predicts query latency based on partition sizes and access frequencies. Quake also dynamically sets query execution parameters to meet recall targets using a novel recall estimation model. Furthermore, Quake utilizes NUMA-aware intra-query parallelism for improved memory bandwidth utilization during search. To evaluate Quake, we prepare a Wikipedia vector search workload and develop a workload generator to create vector search workloads with configurable access patterns. Our evaluation shows that on dynamic workloads, Quake achieves query latency reductions of $1.5$–$38\times$ and update latency reductions of $4.5$–$126\times$ compared to state-of-the-art indexes such as SVS, DiskANN, HNSW, and SCANN.
\end{abstract}

\section{Introduction}
Vector search, the task of finding the $k$-nearest neighbors (KNN) of a query vector against a database of high-dimensional vectors, is fundamental to modern machine learning based search~\cite{grbovic2018real, haldar2019applying, hashemi2021neural, qin2021mixer} and recommendation systems~\cite{okura2017embedding, liu2017related, wang2018billion, pal2020pinnersage, liu2022monolith}. In these applications, a vector represents an item in a metric space, and the distance between vectors reflects semantic similarity. However, performing exact KNN search becomes computationally infeasible on large datasets due to the high dimensionality and volume of data.

To address this challenge, practitioners use approximate nearest neighbor (ANN) indexes, which trade off a controlled amount of search accuracy (recall) for significant reductions in latency. Among these, two broad classes dominate in practice: graph-based and partitioned indexes, each with distinct performance characteristics under dynamic workloads.

% Among these indexes, there are two widely-used types: graph-based indexes and partitioned indexes.

Maintaining low latency, high recall vector search under \textbf{dynamic and skewed workloads} remains a significant challenge for existing indexes. Real-world applications often exhibit non-uniform query distributions and evolving data. For example, in an example Wikipedia search application, popular pages like \textit{Lionel Messi} or \textit{LeBron James} receive disproportionately more queries, resulting in \emph{skewed read patterns}. Additionally, pages are frequently added, updated, or deleted, causing \emph{skewed update patterns} that change over time ~\cite{baranchuk_dedrift_2023}. These factors degrade the performance of existing indexes, leading to increased query latency and reduced recall.

\emph{Graph-based indexes}, such as HNSW~\cite{10.1109/TPAMI.2018.2889473}, DiskANN ~\cite{10.5555/3454287.3455520, singh2021freshdiskann}, and SVS ~\cite{aguerrebere2024locally} construct a proximity graph where each node (vector) is connected to its approximate neighbors. Queries traverse the graph to find approximate nearest neighbors, typically achieving high recall with low latency. However, these indexes face challenges with dynamic workloads because updating the graph structure to accommodate frequent insertions and deletions is computationally intensive~\cite{xu_spfresh_2023}, due to the random access patterns involved in graph traversal and modification.

\emph{Partitioned indexes}, such as SCANN \cite{guo_accelerating_2020, soar_2023}, SPANN \cite{xu_spfresh_2023, chen_spann_nodate}, and  Faiss-IVF \cite{douze2024faisslibrary}, partition the vectors using a clustering algorithm (e.g k-means). Queries are processed by scanning a subset of partitions, balancing recall and latency by adjusting the number of partitions scanned (denoted as \emph{nprobe}).  While attractive due to their simplicity, partitioned indexes face a significant search latency gap when compared with graph indexes. For example, on the \textsc{MSTuring10M} benchmark \cite{bigann}, we found Faiss-IVF takes 44ms per search query while Faiss-HNSW takes only 6.8ms. 
On the other hand, supporting updates in partitioned indexes is less expensive than for graph indexes, as the index structure needs minimal modification when adding or removing vectors. But, existing approaches struggle with dynamic and skewed workloads because they do not adapt to changing access patterns, leading to \emph{imbalanced} partitions that degrade query latency. Recent work has been proposed to resolve imbalances in dynamic workloads by splitting and reclustering imbalanced partitions \cite{xu_spfresh_2023, baranchuk_dedrift_2023}, however, we find these methods degrade recall as nprobe needs to change as the index structure changes.

In this work, we study the problem of minimizing query latency to meet a fixed recall target for dynamic vector search workloads, where both the queries and the base vectors can change over time. To address this problem, we develop \textbf{Quake}, a partitioned index for ANN search that minimizes query latency by adapting the index structure to the workload. \edit{Quake’s two primary algorithmic contributions are:}

First, Quake employs an \textbf{adaptive hierarchical partitioning} scheme that modifies the partitioning by minimizing the cost (derived from a proposed \textbf{cost model}) of a query. The cost model tracks partition sizes and access frequencies as the workload is processed and determines which partitions are most negatively contributing to overall query latency. Once identified, we split or merge these partitions based on expected cost reduction derived from our proposed cost model. We also demonstrate that our maintenance procedure is stable and converges to a local minimum of the cost model.

Second, we design an \textbf{adaptive partition scanning} scheme that adjusts the number of partitions scanned on-the-fly to meet recall targets for individual queries. We do this by maintaining a recall estimate during query processing based on A) the geometry of the partitioning and B) intermediate results of the query, and once the estimate exceeds the recall target, query processing terminates and the results are returned.

\edit{Furthermore, Quake utilizes NUMA-aware parallelism to maximize memory bandwidth usage on multi-core machines.}

It is a significant challenge to evaluate indexing approaches due to the lack of availability of benchmarks for online vector search. To address this challenge and comprehensively evaluate our approach, we A) prepare a \textbf{Wikipedia vector search workload} derived from publicly available query and update patterns of Wikipedia pages and B) develop a \textbf{workload generator} for creating workloads with configurable query and update patterns. We will publicly release the Wikipedia workload and workload generator as evaluation tools for the community to use. Using these, we conduct a comprehensive evaluation of Quake in comparison to seven baseline approaches.

\begin{enumerate}[leftmargin=*, itemsep=-1pt, topsep=0pt]
    \item Quake achieves the lowest search time across all dynamic workloads compared to state-of-the-art graph indexes, with $1.5$-$13\times$ lower search latency than HNSW, DiskANN, and SVS while having $18$-$126\times$ lower update latency.
    \item We also find that APS matches the nprobe of an oracle across recall targets on \textsc{Sift1M}, with only a $17$-$29\%$ increase in latency relative to the oracle. 
    \item APS performs on-par or better than existing early termination methods~\cite{zhang2023fast, chen_spann_nodate, li2020improving} and requires no offline tuning. %Additionally, in contrast to existing methods for early termination, APS requires no offline tuning.
    \item Quake's NUMA-aware query processing exhibits linear scalability and high memory bandwidth utilization on the \textsc{MSTuring100M} dataset. Quake achieves $20\times$ and $4\times$ lower query latency compared to single-threaded and non-NUMA aware configurations, respectively.% version and $4\times$ lower latency compared to a non-NUMA aware configuration.
\end{enumerate}

\section{Motivation and Challenges}
\label{sec:motivation}

Efficient vector search is critical for large-scale systems used in recommendation, semantic search, and information retrieval. These applications demand the ability to process a high volume of nearest neighbor queries with low latency, even as the underlying data evolves. To meet these requirements, vector databases—such as Milvus \cite{wang2021milvus}, Pinecone \cite{pinecone-url}, AnalyticDB-V \cite{wei2020analyticdb}, VBASE \cite{zhang2023vbase}, and Qdrant \cite{noauthor_qdrant_nodate}—utilize specialized vector indexes (e.g., Faiss-IVF, HNSW, Vamana) that support fast approximate nearest neighbor (ANN) queries. However, serving these dynamic workloads introduces significant challenges in maintaining query performance and accuracy as data and query patterns shift over time.

\subsection{Vector Search Workload}
A vector search workload is a continuous, evolving stream of \textbf{queries} and \textbf{updates}:
\begin{itemize}[leftmargin=*, itemsep=0pt, topsep=0pt]
    \item \textbf{Queries:} Given a query vector $q$, the goal is to find the top-$k$ nearest neighbors in a set $\mathbf{X}$. Exact linear search is too slow for large, high-dimensional datasets, so ANN indexes are used. These indexes approximate nearest neighbors with controlled recall to lower latency by orders of magnitude.
    \item \textbf{Updates:} The dataset evolves over time. Insertions add new vectors representing fresh content (e.g., new products, trending news articles), and deletions remove outdated entries. Typically, updates are applied in a batched fashion.
\end{itemize}

\textbf{Recall@k} is the standard metric for accuracy, defined as:
$\frac{|\mathbf{G} \cap \mathbf{R}|}{k}$ where \(\mathbf{R}\) is the vectors returned by the approximate search, and \(\mathbf{G}\) is the ground truth set. Maintaining a consistent recall target (e.g., $>90\%$) and low latency (e.g., milliseconds per query) as both data and query patterns shift is a key challenge. The complexity of these workloads stems from their inherently dynamic and skewed nature, which few existing indexing methods handle gracefully.

\subsection{Why Real-World Workloads are Hard}

\paragraph{Skewed Read Patterns}
In practice, user queries concentrate on popular items. For example, queries against a Wikipedia-derived dataset tend to focus on a small subset of entities at any given time. As a result, certain partitions or graph regions of the index are accessed disproportionately often.

\paragraph{Skewed Write Patterns}
Insertions and deletions are also rarely uniform. New data often arrives in bursts—e.g., new Wikipedia pages added monthly, new products introduced ahead of a shopping season, or newly relevant embeddings generated by continuously updated language models. 

\begin{figure}[t]
    \centering
    \begin{subfigure}[t]{0.48\linewidth}
        \centering
        \includegraphics[width=\linewidth]{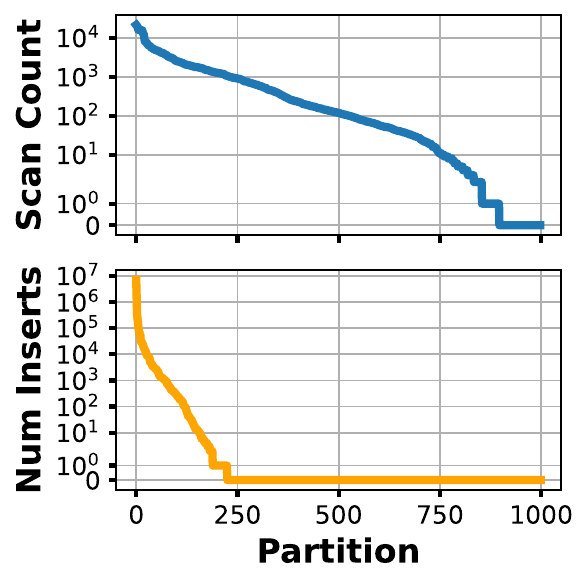}
        \caption{Read (top) and write skew.}
        \label{fig:wiki_combined}
    \end{subfigure}%
    \hfill
    \begin{subfigure}[t]{0.48\linewidth}
        \centering
        \includegraphics[width=\linewidth]{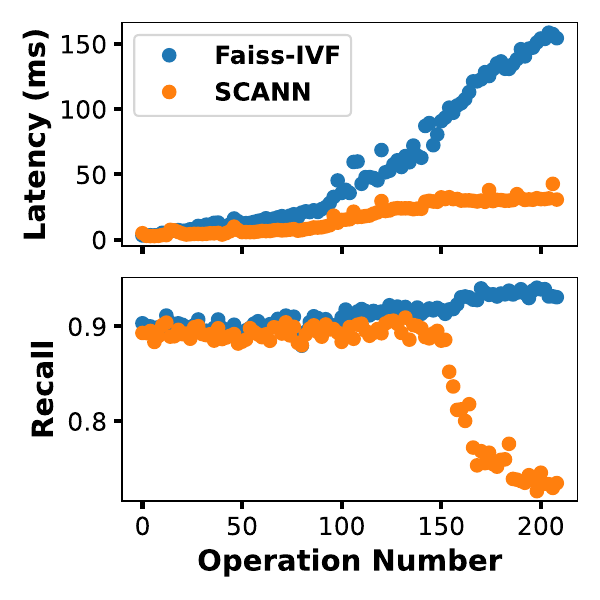}
        \caption{Query performance.}
        \label{fig:ivf_latency}
    \end{subfigure}
    \caption{Skewed access patterns of Faiss-IVF index partitions in the \textsc{Wikipedia-12M} workload and their effect on query performance for Faiss-IVF and SCANN}
    \label{fig:combined_figures}
\end{figure}

\paragraph{Real-World Example: Wikipedia-12M} In our evaluation, we prepared \textsc{Wikipedia-12M}, a workload based on a subset of Wikipedia articles derived from publicly available monthly pageview statistics \cite{wiki}. Over 103 months, the dataset grows from millions to tens of millions of vectors. Popular articles dominate query traffic, while embeddings of newly created pages accumulate in certain regions of the embedding space. This workload shows read skew and write skew, as evidenced by Figure \ref{fig:wiki_combined}, reads and writes predominantly affect a small portion of the index. 

\begin{table}[t]
\small
\centering
\caption{\edit{Comparison of updatable vector indexes.} \textbf{Tuning}: Requires manual parameter tuning in indexing/query processing. \textbf{Maintenance}: Modifies index with incremental updates. \textbf{Adaptive}: Uses query information to inform indexing.}
\label{tab:method_comparison}
\begin{tabular}{lccc}
\toprule
\textbf{Method} & \textbf{Tuning} & \textbf{Maint.} & \textbf{Adaptive} \\
\midrule
Quake \textbf{(Ours)} & \xmark & \cmark & \cmark \\
Faiss-IVF~\cite{douze2024faisslibrary} & \cmark & \xmark & \xmark \\
DeDrift~\cite{baranchuk_dedrift_2023} & \cmark & \cmark & \xmark \\
SpFresh~\cite{xu_spfresh_2023} & \cmark & \cmark & \xmark \\
SCANN~\cite{guo_accelerating_2020, soar_2023} & \cmark & \cmark & \xmark \\
DiskANN~\cite{ni_diskann_2023, singh2021freshdiskann} & \cmark & \cmark & \xmark \\
Faiss-HNSW~\cite{10.1109/TPAMI.2018.2889473} & \cmark & \xmark & \xmark \\
SVS~\cite{aguerrebere2024locally} & \cmark & \cmark & \xmark \\
\bottomrule
\end{tabular}
\end{table}

\subsection{Shortcomings of Existing Approaches}
Existing indexes were often developed and evaluated under assumptions of static data distributions; conditions not met in real-world use cases. Table~\ref{tab:method_comparison} compares a range of state-of-the-art vector indexes. Although widely adopted in vector databases, none fully solve the problem of maintaining low-latency, high-recall search under dynamic, skewed workloads without constant manual intervention or offline tuning.

\paragraph{Graph Indexes}  
Graph-based index systems, such as HNSW~\cite{10.1109/TPAMI.2018.2889473}, DiskANN~\cite{ni_diskann_2023, singh2021freshdiskann}, and SVS ~\cite{aguerrebere2024locally} construct a proximity graph where each node represents a vector connected to its approximate neighbors. These indexes achieve high recall with low latency in static settings by efficiently traversing the graph to locate nearest neighbors using a process known as \emph{greedy traversal}. However, maintaining the graph structure under frequent updates is computationally intensive, as each update may require rewiring multiple edges to preserve graph connectivity and proximity properties. Our evaluation (Table \ref{tab:method_comparison}) shows that update latency can be multiple orders of magnitude higher than partitioned indexes.

\paragraph{Partitioned Indexes}  
Partitioned indexes such as Faiss-IVF~\cite{douze2024faisslibrary}, SCANN~\cite{guo_accelerating_2020, soar_2023}, and SpFresh~\cite{xu_spfresh_2023} divide the vector space into disjoint partitions using a clustering algorithm such as k-means. Queries are processed by scanning a subset of partitions to retrieve approximate nearest neighbors. Partitioned indexes are more update-friendly than graph-based method since insertions and deletions leads to sequential access. For write skewed workloads some partitions become significantly larger, degrading query latency, this can be exacerbated by read skew if large partitions are also more frequently accessed ("hot partitions"). Query processing is \emph{memory-bound}, as achieving high recall requires scanning many megabytes of data across multiple partitions, for example reaching a recall target of 90\% on the \textsc{MSTuring100M} dataset requires each query to scan 1GB of vectors. Moreover, most partitioned indexes use a fixed number of partitions to probe (\emph{nprobe}), which does not adapt to changing data distributions or query patterns, leading to either insufficient recall or excessive data scanning. The challenges yield subpar performance for partitioned indexes on real-world workloads. For example, Figure \ref{fig:ivf_latency} shows the degradation of latency and recall over time when using Faiss-IVF and SCANN with a fixed nprobe on \textsc{Wikipedia-12M} (workload details in Section~\ref{sec:experiments}).

\edit{\paragraph{Early Termination}}
Early-termination methods have been proposed to reduce query latency or meet recall targets in partitioned indexes by dynamically adjusting the number of partitions scanned per query. \textbf{SPANN}~\cite{chen_spann_nodate} applies a simple rule: it prunes partitions once the centroid distance exceeds a user-tuned threshold relative to the closest centroid. \textbf{LAET}~\cite{li2020improving} is a learning-based approach that predicts the required \emph{nprobe} per query using a trained model, but still requires dataset-specific training and calibration for each recall target. \textbf{Auncel}~\cite{zhang2023fast} uses a geometric model to estimate when recall for a given query, setting \emph{nprobe} per query, but its conservative estimation leads to substantial overshooting of the recall target (See Figure 13 in \cite{zhang2023fast}). All three methods require tuning or calibration and do not adapt to changes in the index structure or data distribution.

\subsection{Technical Challenges for Partitioned Indexes}
The following technical challenges are yet to be solved by existing partitioned indexes
\begin{enumerate}[leftmargin=*, itemsep=-.25em, topsep=0pt]
    \item \textbf{Adaptation to Queries} Query adaptivity is overlooked by existing partitioned index approaches and exhibits an opportunity for optimization, particularly for maintaining hot partitions induced by read skew.
    \item \textbf{Online Adjustment of Nprobe} As the index structure and data change, partitioned indexes need to adjust the number of partitions scanned or recall will suffer. \edit{Existing early termination works are insufficient as they assume a static index and require retuning as the index and data change.}
    \item \textbf{Performance Gap with Graph Indexes} Standard partitioned indexes such as Faiss-IVF are memory bound, and exhibit an order of magnitude higher query latency in comparison than graph indexes.
\end{enumerate}

Quake is our solution to these technical challenges. Quake A) adapts the index structure to queries by utilizing maintenance that minimizes a cost model for query latency, B) using a recall estimation model, Quake individually sets nprobe for queries to meet recall targets as the index structure changes, and C) uses NUMA-aware parallelism in order to saturate memory bandwidth during query processing, closing the performance gap with graph indexes. We next briefly discuss other related work before covering Quake in detail.

\section{Solution Overview}
\label{sec:overview}

\begin{figure}[h]
    \centering
    \includegraphics[width=0.5\textwidth]{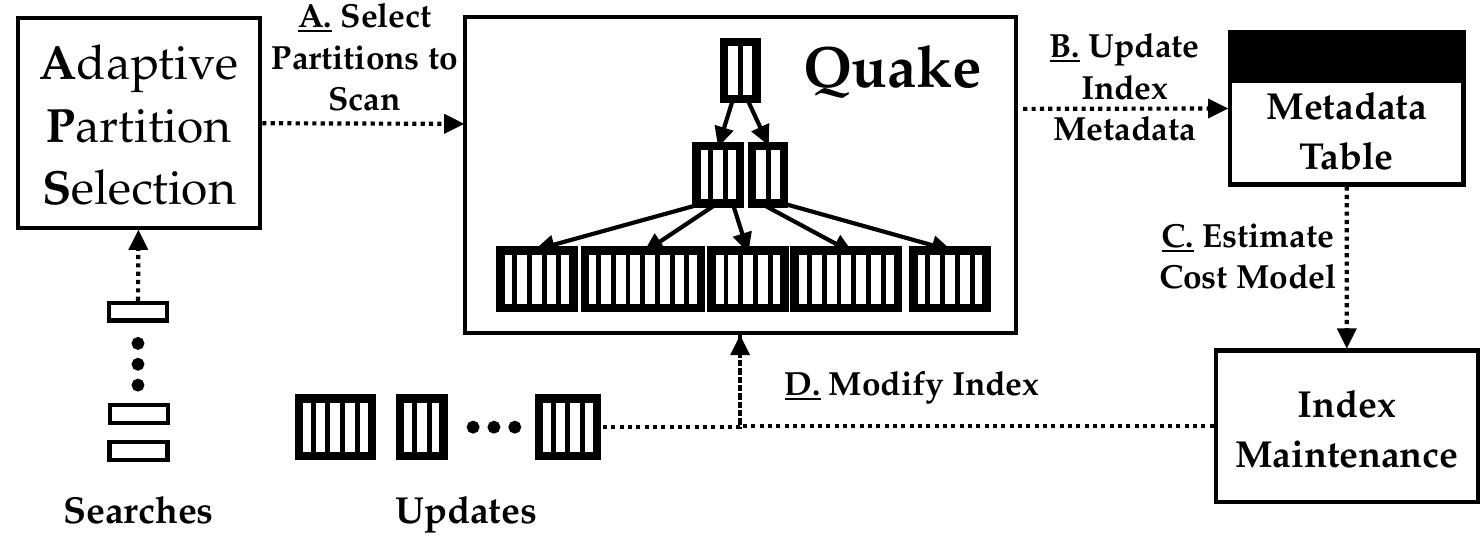}
\caption{Quake Architecture Diagram. Search queries use Adaptive Partition Selection (APS) to determine which partitions to scan (\textbf{A}). Scanning partitions modifies access patterns of the index, tracked in the metadata table (\textbf{B}). A cost model is used to determine which maintenance actions to take (\textbf{C}) where the chosen maintenance actions modify the index (\textbf{D}). This process operates in a continuous online fashion as search and update queries (inserts/deletes) are issued to the index. }
    \label{fig:quake_overview}
\end{figure}

\paragraph{Index Structure} Quake organizes the vectors in a multi-level index, where each level is a partitioned index similar to Faiss-IVF~\cite{douze2024faisslibrary}. The lowest level in the index is constructed by organizing the vectors into disjoint partitions (using k-means clustering) where each partition has a representative centroid. These centroids can be further partitioned in a similar manner to create additional levels in the index. Search queries scan the index in a top-down fashion, finding the nearest centroids at each level to determine the partitions to scan in the next level. Partitions in the lowest level contain the actual vectors and subsets of these partitions are scanned to return the k-nearest neighbors. Utilizing a multi-level design enables us to employ fine-grained partitioning of vectors at large scale (shown to improve search quality~\cite{chen_spann_nodate}), while mitigating the high cost of scanning centroids.
% mitigates the cost of scanning centroids, allowing us to use fine-grained partitioning of vectors at large scale, which prior work has shown to be beneficial \cite{chen_spann_nodate}.

\paragraph{Adaptive Incremental Maintenance} Inserts and deletes modify the Quake data structure by appending vectors to and removing vectors from index partitions. Insertions traverse the index structure top-down to find the nearest partition in the lowest level to the inserted vector and append to that partition. Deletes use a map to find the partition containing the vector to be deleted and the vector is removed from the partition with immediate compaction. As demonstrated in Section \ref{sec:motivation}, modifications can negatively affect index performance over time, requiring maintenance (Figure \ref{fig:ivf_latency}).
Quake uses the following maintenance actions in order to minimize query latency:

\begin{enumerate}[leftmargin=*, itemsep=0em, topsep=0pt]
\item \textbf{Split Partition:} Uses k-means clustering to split a partition into two, removing the old partition and its centroid and adding two new partitions and centroids. To mitigate potential overlap due to the new partitions, we perform additional iterations of k-means clustering over the partitions neighboring the split partitions (by centroid distance).
\item \textbf{Merge Partition:} Removes a partition and its centroid, reassigning the vectors of the removed partition to the remaining partitions in the index.
\item \textbf{Add Level:} Adds a level of partitioning to the index by partitioning the current top-level using k-means clustering.
\item \textbf{Remove Level:} Removes current top-level and merges the partitions in the next level.
\end{enumerate}

Quake uses a cost model that estimates query latency to determine if maintenance actions should be taken and which partitions to apply them to. The cost model is a function of partition access patterns and sizes to determine which partitions are contributing most to the overall query latency. We check for maintenance after each operation by evaluating the cost model, but the maintenance frequency is configurable. Partitions with the largest cost contribution are considered for split or deletion. Intuitively, frequently accessed and/or large partitions are split and infrequently accessed and/or small partitions are merged as they do not justify the overhead of maintaining a centroid. See Section \ref{sec:maintenance} for details on the cost model and maintenance methodology.

\paragraph{Adaptive Partition Scanning} In order to determine the number of partitions a search query should scan to reach a given recall target, we apply \emph{Adaptive Partition Scanning} (APS) at each level of the index. APS solves a critical problem for partitioned indexes when applied to dynamic workloads: as the number and contents of partitions change, the number of partitions scanned needs to change, otherwise recall will degrade (Figure \ref{fig:ivf_latency}). APS maintains a recall estimator based on the intermediate top-k results of the query and the geometry of neighboring partitions. As more partitions are scanned, the intermediate results and recall model are updated and when the recall estimate exceeds the target recall, the results are returned. To mitigate overheads introduced by the estimator, we use pre-computation of expensive geometric functions and only update the estimate when the intermediate results have changed significantly. APS supports both Euclidean and inner-product distance metrics. We cover APS in Section \ref{sec:adaptive_partition_scanning}.

\paragraph{NUMA-Aware Query Processing} Modern multi-core servers often use Non-Uniform Memory Access (NUMA) architectures, where memory close to a processor’s local node is faster to access than remote memory. Quake is designed to capitalize on this heterogeneous memory. It distributes index partitions across NUMA nodes. To minimize remote memory access, Quake employs affinity-based scheduling, and supports work stealing within a NUMA node to mitigate workload imbalances. By co-locating computation with the relevant data, Quake reduces remote memory accesses, saturates memory bandwidth, and thus lowers query latency. See Section~\ref{sec:numa_aware_query_processing} for details on Quake’s NUMA-aware optimizations.

% to minimize data movement

\section{\edit{Adaptive Incremental Maintenance}}
\label{sec:maintenance}

We present our adaptive maintenance methodology, beginning with a cost model that estimates each partition’s contribution to query latency and guides maintenance decisions. Next, we describe the available maintenance actions, analyzing their impact on the cost model. We then detail the multi-stage decision workflow that prioritizes beneficial actions and conclude with a concrete example.

\subsection{Cost Model}

The cost model estimates the query latency contributed by each partition in the index. Estimating the per-partition latency contribution enables targeted maintenance to the partitions most affecting query performance.

\paragraph{Partition Properties} 
Consider an index with $L$ levels, numbered $l = 0,1,\dots,L-1$. Level $l$ contains $N_l$ partitions. The base level corresponds to $l=0$ and contains partitions of the original dataset vectors. Higher levels contain partitions of centroid vectors that summarize the partitions in the level below. At the top level, $l=L-1$, there is a single partition containing top-level centroids.

Each partition $j$ at level $l$ has a size $s_{lj}$ (the number of vectors it contains) and an access frequency $A_{l,j} \in [0.0, 1.0]$. $A_{l,j}$ denotes the fraction of queries, measured in a sliding window $W$, that scan the partition $j$ at level $l$. The cost model is primarily driven by these sizes and access frequencies.

\paragraph{Partition Cost}
A partition $(l,j)$ contributes latency proportional to its size and how frequently it is accessed. Let $\lambda(s)$ be the latency function for scanning $s$ vectors. We measure $\lambda(s)$ through offline profiling. The cost of partition $(l,j)$ is:
\begin{equation}
C_{lj} = A_{lj} \cdot \lambda(s_{lj})
\label{eq:CostPartition}
\end{equation}

\paragraph{Total Cost}
The overall query latency (cost) estimate is the sum across all levels and partitions
\begin{equation}
C = \sum_{l=0}^{L-1} \sum_{j=0}^{N_l-1} A_{lj} \cdot \lambda(s_{lj})
\label{eq:CostTotal}
\end{equation}

\paragraph{Interpretation} The cost model reflects the relationship between partition size, access frequency, and query latency. The fundamental trade-off that needs to be balanced is the number and size of partitions. Larger partitions require more time to scan, increasing latency, but reducing the total number of partitions and the overhead of scanning centroids. Conversely, smaller, fine-grained partitions reduce the number of vectors needed to scan to reach a high recall but increase the overhead of scanning centroids. The model further captures that frequently accessed partitions dominate the total cost, motivating targeted maintenance actions to balance these trade-offs.

\paragraph{Guiding Maintenance Decisions} Maintenance actions such as splitting or deleting aim to reduce the total cost \(C\). Each action is evaluated based on its predicted change in cost:
\begin{equation}
\Delta C = C_{\text{after}} - C_{\text{before}}
\label{eq:DeltaC}
\end{equation}
where \(C_{\text{before}}\) and \(C_{\text{after}}\) are the total costs before and after the action, respectively. Actions are applied only if \(\Delta C < -\tau\), where $\tau$ is a non-negative tunable threshold, ensuring monotonic improvement in query performance. By focusing on reducing \(C\), the index is dynamically restructured to maintain efficient query performance under varying workloads.

\subsection{Conducting Maintenance}
\label{sec:maintenance_actions}

Maintenance at each level of the index proceeds in three phases—\emph{estimate}, \emph{verify},
and \emph{commit / reject}.  We first list the available actions,
then derive their cost deltas, describe the workflow, and finally walk
through a concrete example.

\subsubsection{Maintenance Actions}

To minimize query latency, Quake employs a series of maintenance actions that dynamically adjust the index structure in response to evolving workloads. Here we define the maintenance actions and then analyze the impact of each maintenance action on the overall cost model. 

\paragraph{Split Partition} If a partition $(l,j)$ is too large or frequently accessed, we consider splitting it into two partitions $(l,j_L)$ and $(l,j_R)$. We apply $k$-means clustering within that partition, forming two smaller partitions with their own centroids. The original partition is removed and its vectors are reassigned. A subsequent \emph{partition refinement} step adjusts vector assignments to ensure minimal overlap and balanced partition sizes.

\paragraph{Partition Refinement} After a split, refinement uses $k$-means (seeded by current centroids) on nearby partitions to mitigate overlap and ensure that each vector is assigned to its most representative partition. Nearby partitions are determined by finding the $r_f$ nearest centroids to the split centroids, where $r_f$ is a tunable parameter (typically between 10 and 100). This is a generalization of the reassignment procedure used in SpFresh/LIRE \cite{xu_spfresh_2023}, using additional rounds of k-means prior to reassignment, and has been applied successfully by recent index maintenance works \cite{mageirakos2025cracking} and \cite{mohoney2024incremental}. Refinement avoids performance degradation by mitigating overlap and ensuring vectors are assigned to their most representative partition.

\paragraph{Merge Partition} If a partition is rarely accessed and below a minimum size threshold, we consider deleting it to remove the cost of maintaining its centroid. After deletion, the vectors are reassigned to their respective nearest existing partitions. This can reduce total cost by removing a low-benefit partition, although the reassignment may increase the size (and thus cost) of other partitions, and therefore careful consideration is needed before conducting a merge.

\paragraph{Adding and Removing Levels} If the number of centroids in the top level grows beyond a threshold, we add a new top level by clustering those centroids. Conversely, if the top level becomes too sparse (below a configured lower threshold), we remove the top level and merge its centroids in the level below. Both actions help maintain hierarchy balance and control centroid-scanning overhead. We defer discussion of this to our technical report \cite{mohoney2025quakeadaptiveindexingvector}.

% ------------------------------------------------------------
\subsubsection{Cost Deltas}
\label{sec:maintenance_actions:deltas}

The maintenance loop treats every candidate action as a \emph{proposed
edit} to the index and scores it by the change it would induce in the
total cost (Eq.~\eqref{eq:DeltaC}).

We tentatively accept an action whenever $\Delta C < -\tau$.  Below we give the exact
$\Delta C$ formulas for the primary maintenance actions:
\emph{split} and \emph{merge}.  Full derivations are in the
technical report \cite{mohoney2025quakeadaptiveindexingvector}; here we show only the final expressions.

\paragraph{Exact Split Delta}
Splitting a hot or oversized partition $(l,j)$ into children
$(l,j_L)$ and $(l,j_R)$ inserts one new centroid at the parent level,
changing the overhead by
$\Delta O^{+} = \lambda(N_{l}+1) - \lambda(N_{l})$.  The resulting
cost difference is

\begin{equation}
\begin{aligned}
\Delta\text{Split}_{l,j} \;=\;
  &\underbrace{\Delta O^{+}}_{\text{new centroid}}
  - A_{l,j}\,\lambda(s_{l,j})  \\
  &+ A_{l,j_L}\,\lambda(s_{l,j_L})
  + A_{l,j_R}\,\lambda(s_{l,j_R})
\end{aligned}
\label{eq:delta_split_exact}
\end{equation}

where the first term pays for the extra centroid, the second removes
the old scan cost, and the last two add the costs of scanning the new,
smaller partitions. Note that we do not explicitly model the effect of refinement, as refinement does not change the number of partitions. Its impact is captured automatically as statistics are collected from future queries, so we omit it from the $\Delta$-formula and let later maintenance iterations adjust if necessary.

\paragraph{Exact Merge Delta}
Deleting a cold, tiny partition $(l,j)$ removes a centroid
($\Delta O^{-}= \lambda(N_{l}-1)-\lambda(N_{l})$) and redistributes
its vectors to a receiver set $R_{l,j}$.  Let $\Delta s_{m}$ and
$\Delta A_{m}$ be the resulting size and frequency bumps for each
receiver $m$.  Then

\begin{equation}
\begin{aligned}
\Delta\text{Merge}_{l,j} \;=\;
  &\Delta O^{-} - A_{l,j}\,\lambda(s_{l,j}) \\
  &+ \sum_{m\in R_{l,j}}
     \bigl[(A_{m}{+}\Delta A_{m})\,
           \lambda(s_{m}{+}\Delta s_{m})
           - A_{m}\,\lambda(s_{m})\bigr]
\end{aligned}
\label{eq:delta_merge_exact}
\end{equation}

captures both the benefit of deleting the partition and the
penalty of swelling its neighbors.

\paragraph{Estimating Deltas}
At decision time we do not yet know the post-action quantities
$\{s_{l,j_L},A_{l,j_L},\dots\}$ or the true $\Delta s_{m},
\Delta A_{m}$.  We therefore use a lightweight \emph{estimate}
based on two assumptions: 1) \textbf{Balanced Split: } $s_{l,j_L}\!\approx\!s_{l,j_R}\!\approx\!\tfrac{s_{l,j}}{2}$, and 2) \textbf{Proportional-Access Scaling}: each child inherits a fixed
      fraction~$\alpha$ of the parent’s frequency.

Under these assumptions the split estimate becomes

\begin{equation}
\begin{aligned}
\Delta'\text{Split}_{l,j} \;=\;
  &\Delta O^{+}
  - A_{l,j}\,\lambda(s_{l,j}) \\
  &+ 2\alpha\,A_{l,j}\,
    \lambda\!\bigl(\tfrac{s_{l,j}}{2}\bigr).
\end{aligned}
\label{eq:delta_split_est}
\end{equation}

and the analogous merge estimate, derived with a uniform
redistribution assumption, is located in the technical report \cite{mohoney2025quakeadaptiveindexingvector}.
Immediately after a tentative action we measure the
\emph{actual} sizes (and, for merges, the exact receiving partitions) and
re-evaluate Eqs.~\eqref{eq:delta_split_exact}
or~\eqref{eq:delta_merge_exact}.  If the recomputed gain is still
below $-\tau$ the action is committed; otherwise it is rolled back
(\S\ref{sec:maintenance_actions:workflow}).  This “estimate-then-verify”
strategy is crucial for ensuring
monotonic cost improvement.

\subsubsection{Decision Workflow}
\label{sec:maintenance_actions:workflow}

Maintenance is a \textit{bottom-up} pass over the hierarchy.  
Each level executes the five stages below starting from the base level. This workflow is triggered by the user. An avenue for future work is to develop scheduling policies to call this workflow and limit its scope. In our evaluation, we trigger maintenance after a set amount of queries have been run.

\paragraph{Stage 0: Track Statistics}
At the end of each query batch we update, for every partition
$(l,j)$: (i) size $s_{l,j}$, (ii) access count over the sliding window of queries 
$W$, giving $A_{l,j}= \text{hits}(l,j)/|W|$.
These values are inputs of the cost model.

\paragraph{Stage 1: Estimate}
For the current level $l$ compute the estimate
$\Delta'$ (\S\ref{sec:maintenance_actions:deltas}) of splitting and deleting for every
partition.  Tentatively apply any action with $\Delta' < -\tau$.

\paragraph{Stage 2: Verify}
Immediately after performing a tentative action, we measure the actual resulting partition sizes (and the exact receiver partitions for merges). We recompute the cost delta using these known values but retain the original frequency assumptions from Stage 1.

\paragraph{Stage 3: Commit / Reject}
\[
\Delta<-\tau\;\rightarrow\;\textit{commit},\quad
\Delta\ge-\tau\;\rightarrow\;\textit{reject}.
\]
Rejection discards the action and keeps the previous state of the partition(s), in order to prevent cost increases.

\paragraph{Stage 4: Propagate Upward.}
Repeat Stages 1-3 on the next level $l{+}1$,  

\smallskip
\noindent\textit{Safety:}  
Because every level enforces the same
\mbox{$\Delta<-\tau$} guard, total cost across \emph{all} levels
monotonically decreases and the hierarchy converges to a stable state
under a fixed workload distribution (proof in technical report \cite{mohoney2025quakeadaptiveindexingvector}).

\subsubsection{Example Maintenance Workflow}
Below we walk through the \emph{estimate → verify → commit /
reject} loop for two example partitions and show how an
imbalanced split is automatically rejected to prevent accidental cost increases.

%--- set-up ---------------------------------------------------------
\noindent\textit{Set-up:}\;
Consider partitions $P_1$ and $P_2$ with identical size and access frequency, where both contain
$\;s\!=\!500\;$ vectors and appear in
$\;A\!=\!0.10\;$ of queries. From profiling we observe non-linear \footnote{Scan latency is non-linear w.r.t. size due to top-k sorting overhead.} scan latencies for the following partition sizes: $\lambda(50)=250\,\mu\text{s}$, $\lambda(250)=550\,\mu\text{s}$, $\lambda(450)=1050\,\mu\text{s}$, $\lambda(500)=1200\,\mu\text{s}$ Adding a centroid costs $\Delta O^{+}=60\,\mu\text{s}$.
We use a decision threshold of $\tau=4\,\mu\text{s}$ and $\alpha=.5$

\medskip
\begin{enumerate}[leftmargin=*, nosep]

%--- 1. estimate ----------------------------------------------------
\item \textbf{Estimate}\;
   For $P_1$ and $P_2$ the estimate assumes a balanced $250/250$ split
   and $\alpha=0.5$ traffic per child:
\begin{align*}
C_{\text{before}} &= 0.10 \times 1200 = 120\,\mu\text{s},\\
C_{\text{est}}    &= 0.05 \times (550+550)=55\,\mu\text{s},\\
\Delta'           &= 60 - 120 + 55 = -5\,\mu\text{s}.
\end{align*}
Because $\Delta' < -\tau$, \emph{both $P_1$ and $P_2$ are tentatively
split}.

%--- 2. verify ------------------------------------------------------
\item \textbf{Verify}\;
   After splitting we see that $P_1$ has a $250/250$ split, but $P_2$ comes out $450/50$:
\begin{align*}
C_{\text{verify}}(P_2) &= 0.05 \times (1050+250)=65\,\mu\text{s},\\
\Delta(P_2)            &= 60 - 120 + 65 = +5\,\mu\text{s}.
\end{align*}

%--- 3. commit / rollback ------------------------------------------
\item \textbf{Decide}\;
\begin{itemize}[leftmargin=1.3em,itemsep=2pt]
  \item $P_1$: \emph{commit} because $\Delta=-5\,\mu$s $<-\tau = -4 \mu$s
  \item $P_2$: \emph{reject} because $\Delta=+5\,\mu$s $>-\tau = -4 \mu$s
\end{itemize}
   The verify step therefore blocks an imbalanced split that would
   otherwise \emph{increase} query latency.
\end{enumerate}

\section{\edit{Adaptive Partition Scanning (APS)}}
\label{sec:adaptive_partition_scanning}

Adaptive Partition Scanning (APS) dynamically determines the number of partitions to scan per query to achieve a specified recall target $\tau_R$ with minimal latency. APS adapts to evolving workloads and changing index structures, making it particularly effective in dynamic data settings. We first introduce the geometric model underlying APS, followed by a detailed description of the scanning algorithm, and conclude with key performance optimizations. We apply APS at each level of the index independently. For clarity we focus on the Euclidean distance, we discuss inner product metrics in the technical report \cite{mohoney2025quakeadaptiveindexingvector}.

\paragraph{Geometric Model}
To estimate the probability that each partition contains one of the query’s $k$ nearest neighbors, APS uses a geometric interpretation. Given query $\mathbf{q}$ and $\rho$ the distance to the $k$-th nearest neighbor, consider the hypersphere $\mathcal{B}(\mathbf{q},\rho)$. Under a uniform-density assumption, the fraction of this sphere’s volume intersecting partition $\mathcal{P}_i$ estimates the probability that $\mathcal{P}_i$ holds a nearest neighbor:
\begin{equation}
p_i \;=\;
\frac{\operatorname{Vol}\!\bigl(\mathcal B(\mathbf q,\rho)\cap\mathcal P_i\bigr)}
     {\operatorname{Vol}\!\bigl(\mathcal B(\mathbf q,\rho)\bigr)},
\label{eq:aps-pi}
\end{equation}

Because we do not know the true distance of the $k$-th nearest neighbor apriori, we set $\rho$ to the current $k$-th nearest neighbor observed and update it online as partitions are scanned.

\paragraph{Intersection Volume Approximation}
Exact computation of intersection volumes between a sphere and high-dimensional Voronoi partition boundaries is infeasible, as partitions are intersections of multiple half-spaces. Instead, we approximate each partition as a single half-space, defined by the perpendicular bisector between the query’s nearest centroid $\mathbf{c}_0$ and each neighboring centroid $\mathbf{c}_i$. This simplification results in a hyperspherical cap whose volume $v_i$ has a closed-form expression via the regularized incomplete beta function \cite{li2010concise, lee2014concise} (see technical report \cite{mohoney2025quakeadaptiveindexingvector}).

\paragraph{Nearest Partition Volume Approximation}
The half-space approximation is invalid for the nearest partition $\mathcal{P}_0$, since the query lies within it. Instead, we first compute hyperspherical cap volumes $v_j$ for the remaining $M-1$ candidate partitions and normalize these so that $\sum_{j=1}^{M-1} v_j = 1$. The probability $p_0$ that no neighbor is located outside $\mathcal{P}_0$ is:

\begin{equation}
p_0 = \prod_{j=1}^{M-1}(1 - v_j),
\label{eq:aps-p0}
\end{equation}

with the remaining probability distributed proportionally among other partitions according to their volumes $v_i$:

\begin{equation}
p_i = (1 - p_0)\, v_i.
\label{eq:aps-pi-final}
\end{equation}

\begin{figure}[h]
\centering
\includegraphics[width=0.2\textwidth]{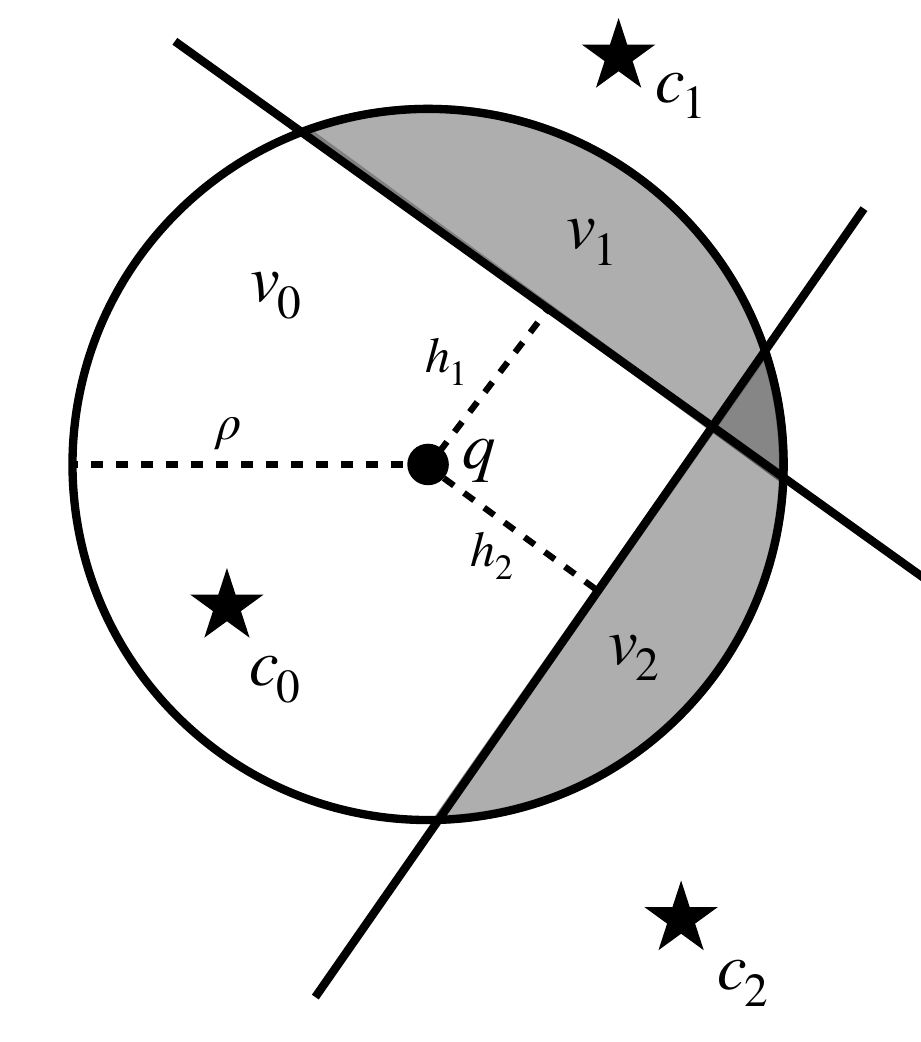}
\caption{The query hypersphere (centered at $\mathbf{q}$ with radius $\rho$) intersecting partition boundaries. The intersection volumes $v_1$ and $v_2$ correspond to the probability of finding a nearest neighbor in partitions $\mathcal{P}_1$ and $\mathcal{P}_2$, respectively.}
\label{fig:geometric_model}
\end{figure}

%-------------------------------------------------------------------------------
\subsection{APS algorithm}
\label{sec:aps-algo}
%-------------------------------------------------------------------------------
Algorithm~\ref{alg:aps} details the APS procedure. Given query $\mathbf{q}$, recall target $\tau_R$, and the initial candidate fraction $f_M$:
\begin{enumerate}[leftmargin=*,nosep]
    \item Scan partition $\mathcal{P}_0$, initializing the query radius $\rho$.
    \item Compute probabilities $p_i$ for each remaining candidate partitions based on radius $\rho$.
    \item Iteratively scan partitions in descending probability order until cumulative recall exceeds target $\tau_R$, updating radius $\rho$ and recomputing probabilities whenever $\rho$ shrinks significantly (beyond threshold $\tau_\rho$).
\end{enumerate}

This process is conducted at each level of the index. To avoid propagating errors from searching higher levels, we fix the recall target to 99\% for the higher levels (Table \ref{tab:two_level_results}).

\begin{algorithm}[h]
\small
\caption{Adaptive Partition Scanning (APS)}
\label{alg:aps}
\begin{algorithmic}[1]
\Require query $\mathbf q$, centroids $\mathcal C$, recall target $\tau_R$,
        initial candidate fraction $f_M$, recompute threshold $\tau_\rho$, $k$
\Ensure  $k$ nearest neighbors of $\mathbf q$
\State $R \leftarrow$ empty max-heap of size $k$
\State $\mathcal S \leftarrow$ $f_M *  N_l$ nearest centroids to $\mathbf q$
\State scan $\mathcal P_0$; update $R$; set $\rho$
\For{each $(\mathbf c_i,\mathcal P_i) \in \mathcal S\!\setminus\!\{\mathbf c_0\}$}
    \State compute $p_i$
\EndFor
\State $r \leftarrow p_0$; $m \leftarrow 1$
\While{$r<\tau_R$ \textbf{and} unscanned candidates remain}
    \State choose $i$ with maximal $p_i$; scan $\mathcal P_i$; update $\mathcal H$
    \State $m \leftarrow m + 1$
    \State $\rho' \leftarrow$ distance to $k$-th in $\mathcal H$
    \If{$|\rho' - \rho| > \tau_\rho \rho$}
        \State $\rho \leftarrow \rho'$; recompute $p_j$
    \EndIf
    \State $r \leftarrow r=\sum^{m-1}_{i=0}p_i$
\EndWhile
\State \Return $R$
\end{algorithmic}
\end{algorithm}

\paragraph{Performance Optimizations}
APS incorporates two optimizations to minimize computational overhead. First, it precomputes values of the regularized incomplete beta function at 1024 evenly spaced points in $[0,1]$ and linearly interpolates during queries. Second, partition probabilities are recomputed only when the query radius $\rho$ shrinks by more than a relative threshold $\tau_{\rho}$. Table~\ref{tab:recall_overhead} shows these optimizations reduce query latency by 29\% on \textsc{Sift1M} without sacrificing recall.

\begin{table}[t]
\small
\centering
\caption{Mean single-threaded query latency and recall for APS variants on \textsc{Sift1M} dataset at recall target 90\%. \textbf{APS-RP:} recomputes probabilities after each partition scan without precomputation. \textbf{APS-R:} recomputes after each partition scan with precomputation. \textbf{APS:} recomputes probabilities only if query radius changes by more than $\tau_{\rho}=1\%$, using precomputed beta function values.}
\label{tab:recall_overhead}
\begin{tabular}{|l|c|c|}
\hline
\textbf{Configuration} & Recall & Search Latency \\ \hline
APS      & 91.2\%   & .48 ms                      \\ \hline
APS-R    & 91.2\% & .59 ms                         \\ \hline
APS-RP   & 91.2\% & .68 ms                         \\ \hline
\end{tabular}
\end{table}

\section{Quake Implementation}
\label{sec:numa_aware_query_processing}

Here we discuss NUMA-aware query processing and implementation details of Quake.

\paragraph{NUMA Data Placement and Query Processing} Query processing in partitioned vector indexes is memory-bound, and therefore increasing the effective memory bandwidth available to the system will reduce query latency. NUMA-aware intra-query parallelism has been applied in the context of relational database systems to great success \cite{10.1145/2588555.2610507, 10.14778/3015274.3015275}, but has yet to be applied to vector databases. 

In order to maximize memory bandwidth utilization, Quake distributes index partitions across NUMA nodes and ensures that cores only scan partitions resident in their respective node. Quake assigns index partitions to specific NUMA nodes using round-robin assignment. This assignment procedure allows for simple load balancing as partitions are added to the index by the maintenance procedure.

To maximize the benefits of data placement, Quake employs partition affinity and NUMA-aware work scheduling. Partitions are bound to specific CPU cores. This binding ensures that partitions are always scanned by the same core to maximize cache utilization. Queries are scheduled to worker threads based on the location of the data partitions they need to access. When a query requires scanning multiple partitions, the work is divided among threads on the relevant NUMA nodes where the partitions reside. By aligning thread execution with data placement, Quake minimizes remote memory accesses to maximize memory bandwidth utilization.

% Placeholder for Figure
% \begin{figure}[h]
% \centering
% \includegraphics[width=0.8\textwidth]{figures/numa_architecture.png}
% \caption{Illustration of NUMA-aware data placement and query processing in Quake.}
% \label{fig:numa_architecture}
% \end{figure}

\paragraph{NUMA-Aware Query Execution with APS} Quake integrates NUMA-aware processing with Adaptive Partition Selection (APS) to dynamically select which partitions to scan based on query requirements and desired recall. The query processing involves both worker threads scanning local partitions and a main thread coordinating the process.

\begin{algorithm}[h]
\small
\caption{NUMA-Aware Query Processing with Adaptive Partition Selection}
\label{alg:numa_aps_query_processing}
\begin{algorithmic}[1]
\Require Query vector $q$, Index partitions ${P_i}$ with locations ${Node_j}$, Recall threshold $\tau_R$, Period to check recall $T_{wait}$
\Ensure Top-$k$ nearest neighbors to $q$ satisfying recall threshold $\tau$
\State \textbf{Initialize}: $R \leftarrow \emptyset$ (global result set), $S \leftarrow$ sorted list of partitions based on distance to $q$ (obtained from searching parent)
\State \textbf{Distribute} $q$ to local memory of NUMA nodes
\ForAll{NUMA nodes $Node_j$ \textbf{in parallel}}
\State $W_j \leftarrow$ worker threads on $Node_j$
\State $P_j \leftarrow$ partitions on $Node_j$ from $S$
\State \textbf{Enqueue} partitions $P_j$ to local job queue
\EndFor
\While{not \textbf{all} partitions in $S$ have been processed}
\State \textbf{Main Thread}:
\State \textbf{Wait} for a predefined interval $T_{wait}$
\State \textbf{Merge} partial results from worker threads into $R$
\State \textbf{Estimate} current recall $r$ using Eqn. \ref{eq:aps-pi}
\If{$r \geq \tau_R$}
\State \textbf{Break} and terminate worker threads
\EndIf
\EndWhile
\State \textbf{Return} top-$k$ results from $R$
\Function{WorkerThread}{$q$, Local Job Queue}
\While{Job Queue not empty}
\State $P_i \leftarrow$ \textbf{Dequeue} next partition from Job Queue
\State Compute distances between $q$ and vectors in $P_i$
\State Update local partial results $R_j$
\State \textbf{Signal} Main Thread about new partial results
\EndWhile
\EndFunction
\end{algorithmic}
\end{algorithm}

\paragraph{Algorithm Explanation:} In Algorithm~\ref{alg:numa_aps_query_processing}, the main steps are:

\begin{enumerate}[leftmargin=*, itemsep=-.25em, topsep=0pt]
\item \textbf{Initialization}: The query vector $q$ is distributed to the local memory of NUMA nodes with relevant partitions. Partitions are sorted using their centroid distance to $q$.

\item \textbf{Worker Threads Execution}: Each NUMA node has worker threads that process partitions assigned to that node. They compute distances between $q$ and vectors in their local partitions, updating their local partial results.

\item \textbf{Main Thread Coordination}: The main thread periodically merges partial results from all worker threads. It uses the APS recall model to estimate the current recall based on the results accumulated so far.

\item \textbf{Adaptive Termination}: If the estimated recall meets or exceeds the threshold $\tau_R$, the main thread returns the top-$k$ results and signals the worker threads to terminate the processing of remaining partitions.

\end{enumerate}

This adaptive approach ensures that the system processes only as much data as needed to meet the recall requirements, improving efficiency and reducing query latency. 

\paragraph{Implementation Details}
We implemented Quake in 7,500 lines of C++ and provide a Python API for ease-of-use. We used primitives in Faiss \cite{douze2024faisslibrary}, PyTorch \cite{libtorch}, and SimSIMD \cite{simsimd} to enable high-performance management of inverted lists, efficient batch tensor operations, and AVX512 intrinsics for fast distance comparisons. We also used a high performance concurrent queue \cite{moody} to prevent contention during coordination of query processing. In addition, we developed a workload generator and evaluation framework in Python to create and evaluate vector search workloads. Quake is open-sourced at \href{https://github.com/marius-team/quake}{https://github.com/marius-team/quake}.

% \subsection{Design Considerations and Future Work}

% While the current NUMA-aware implementation in Quake provides substantial performance gains by saturation memory bandwidth (see evaluation), ongoing development focuses on further optimizations:

% \begin{itemize}
% \item \textbf{Dynamic Data Placement}: Future enhancements may include dynamically adjusting data placement based on real-time access patterns to further balance the workload across NUMA regions.
% \item \textbf{Partition Replication}: For extremely hot partitions, replicating these partitions across NUMA regions could reduce latency at the cost of increased memory usage.
% \item \textbf{Distributed Query Processing}: Our work can be directly applied to the distributed setting, where instead of distributing only across NUMA regions, partitions are distributed across the memory of multiple machines. Supporting distribution allows for increasing the effective memory bandwidth available for query processing, and thereby will decrease query latency.
% \end{itemize}

% These considerations can refine and extend Quake’s NUMA-aware processing to achieve even higher levels of performance and efficiency, especially under imbalanced workloads. We leave these for future work.

\section{Experiments}
\label{sec:experiments}

We evaluate Quake using a number of benchmarks and summarize our main findings:

\begin{enumerate}[leftmargin=*, itemsep=-1pt, topsep=0pt]
    \item Quake achieves the lowest search time across all dynamic workloads compared to state-of-the-art graph indexes, with $1.5$-$13\times$ lower search latency than HNSW, DiskANN, and SVS while having $18$-$126\times$ lower update latency.
    \item We also find that APS matches the nprobe of an oracle across recall targets on \textsc{Sift1M}, with only a $17$-$29\%$ increase in latency relative to the oracle. 
    \item APS performs on-par or better than existing early termination methods~\cite{zhang2023fast, chen_spann_nodate, li2020improving} and requires no offline tuning. %Additionally, in contrast to existing methods for early termination, APS requires no offline tuning.
    \item Quake's NUMA-aware query processing exhibits linear scalability and high memory bandwidth utilization on the \textsc{MSTuring100M} dataset. Quake achieves $20\times$ and $4\times$ lower query latency compared to single-threaded and non-NUMA aware configurations, respectively.% version and $4\times$ lower latency compared to a non-NUMA aware configuration.
\end{enumerate}

\subsection{Workloads}

We performed our evaluation on a diverse set of real-world and synthetic workloads.

\paragraph{Wikipedia-12M} 
This dataset and workload trace are derived from monthly Wikipedia page additions and page-view \cite{wiki} frequencies between April 2013 and December 2021. We consider only pages about people or those linking to people. The dataset begins with 1.6 million pages and grows to 12 million after 103 updates, and therefore the average update size is $\approx$ 100,000 vectors. Embeddings are generated by training DistMult \cite{distmult} graph embeddings (via Marius \cite{mohoney2021marius, waleffe2023mariusgnn}) on the Wikipedia link structure, and use the inner product metric.

The workload simulates monthly inserts of new pages, followed by 100,000 search queries sampling page embeddings with probability proportional to their page views corresponding to roughly a 50/50 read/write ratio. This setting imitates evolving interest and periodic growth of the dataset.

\paragraph{OpenImages-13M}
Using the methodology described by SVS~\cite{aguerrebere2024locally}, we generate a workload of 13M images from the Open Images dataset \cite{Kuznetsova_2020}. Embeddings are produced using Clip \cite{radford2021learningtransferablevisualmodels} in an inner product metric space. The workload maintains a sliding window of 2 million resident vectors and inserts and deletes vectors based on class labels until all 13 million vectors have been indexed at least once. Each insert and delete affects roughly 110K vectors. After each insert and delete operation, we run 1,000 queries randomly sampled from the entire vector set. This scenario stresses both insertion and deletion performance as well as sustained query latency.

\paragraph{Workload Generator} 
To test performance under varying workload properties, we employ a configurable workload generator applicable to any vector dataset. The key parameters are: number of vectors per operation, operation count, operation mix (read/write ratio), and spatial skew. For skewed workloads, vectors are clustered and sampled from to produce queries and updates, reflecting hot spots in the vector space.

We construct two example workloads from a 10M vector subset of the MSTuring~\cite{bigann} dataset using L2 distance:

\begin{itemize}[leftmargin=*, itemsep=-1ex, topsep=0pt]
    \item \textbf{MSTuring-RO}: A pure search workload. We uniformly sample from 100,000 provided query vectors and execute 100 search operations, each querying 10,000 vectors. This setup tests search efficiency in a static setting.
    \item \textbf{MSTuring-IH}: A dynamic workload interleaving inserts and searches. Beginning with 1 million vectors, the dataset grows to 10 million as we process 1,000 operations with a 90\% insert and 10\% search ratio. This tests the ability to handle large-scale growth while maintaining query quality.
\end{itemize}

We use the datasets \textsc{Sift1M} \cite{jegou2010product}, \textsc{Sift10M},  and \textsc{MSTuring100M} \cite{bigann} to conduct microbenchmarks.

\subsection{Experimental Setup}

\edit{Large-scale experiments are run on a 4-socket server with Intel Xeon Gold 6148 CPUs (80 cores, 160 threads), 500 GB RAM across 4 NUMA nodes, and ~300 GB/s total memory bandwidth. Some microbenchmarks (Tables \ref{tab:recall_overhead}, \ref{tab:early_term}, \ref{tab:maintenance_ablation}, and \ref{tab:two_level_results}) are run on a 2023 MacBook Pro with a M2 Max chip.} 

Search queries are processed one at a time and we report the total time to process all queries to reach a target of 90\% recall for $k=100$. \edit{Unless otherwise stated, all search numbers
use a single worker thread. Quake additionally reports a
multi-thread configuration \textbf{Quake-MT} (16 threads) where partition scans are parallelized for individual queries, while \textbf{Quake-ST} uses a single thread for search}. For updates we report the total update and maintenance time, where both Quake and the baselines process updates in batches \edit{using 16 threads}. This setup simulates an online environment where queries arrive individually, and updates are applied in batches. We report maintenance time separately from update latency, as maintenance can be conducted in the background in online systems ~\cite{xu_spfresh_2023}. 

\paragraph{Baselines:} We compare Quake against several state-of-the-art methods, including both partitioned and graph-based indexes:

\begin{itemize}[leftmargin=*, itemsep=-.25em, topsep=0pt]
    \item \textbf{Faiss-IVF}~\cite{douze2024faisslibrary}: A popular inverted file (IVF) index in Faiss. It handles updates but does no maintenance.
    \item \textbf{DeDrift}~\cite{baranchuk_dedrift_2023}: An incremental maintenance strategy designed to reduce clustering drift by periodically reclustering large partitions together with small ones. We implement DeDrift’s logic within Quake.
    \item \textbf{LIRE}~\cite{xu_spfresh_2023}: Maintenance procedure used by SpFresh. LIRE incrementally splits large clusters and deletes small clusters after updates, followed by local reassignments. We implement LIRE’s approach within Quake.
    \item \textbf{ScaNN}~\cite{guo_accelerating_2020}: A state-of-the-art highly optimized partitioned index system. It uses an unpublished incremental maintenance procedure similar to LIRE.
    \item \textbf{Faiss-HNSW}~\cite{10.1109/TPAMI.2018.2889473}: A graph-based approach (HNSW) implemented in Faiss. It supports incremental inserts but not deletes. Thus, for workloads with deletions, we omit Faiss-HNSW from those comparisons.
    \item \textbf{DiskANN}~\cite{singh2021freshdiskann}: System built around the Vamana~\cite{10.5555/3454287.3455520} index with support for dynamic updates.
    \item \textbf{SVS}~\cite{aguerrebere2024locally}: A recently released optimized implementation of the Vamana index with support for dynamic updates. 
\end{itemize}

\begin{table*}[h]
\small
\centering
\begin{tabular}{|l|cccc|cccc|ccc|cccc|}
\hline
\multirow{2}{*}{\textbf{Method}} 
& \multicolumn{4}{c|}{\textsc{Wikipedia-12M}} 
& \multicolumn{4}{c|}{\textsc{OpenImages-13M}} 
& \multicolumn{3}{c|}{\textsc{MSTuring10M-RO}} 
& \multicolumn{4}{c|}{\textsc{MSTuring10M-IH}} \\ \cline{2-16}
& \textbf{S} & \textbf{U} & \textbf{M} & \textbf{T}
& \textbf{S} & \textbf{U} & \textbf{M} & \textbf{T}
& \textbf{S} & \textbf{M} & \textbf{T}
& \textbf{S} & \textbf{U} & \textbf{M} & \textbf{T} \\ \hline
Quake-MT   & 1.53 & .01 & .44 & \textbf{1.98}  &
            .03 & .02 & .10 & \textbf{.15}   &
            .63 & .08 & .71          &
            .54 & .02 & .14 & \textbf{.70} \\ 
Quake-ST   & 9.48 & .01 & .44 & 9.93  &
            .14 & .02 & .10 & .26               &
            2.43 & .08 & 2.51                   &
            2.12 & .02 & .14 & 2.28 \\ 
Faiss-IVF  & 165.8 & .005 & 0 & $165.8^{+}$ &
            .45 & .01 & 0 & .46     &
            12.25 & 0 & 12.25          &
            13.72 & .01 & 0 & 13.73 \\ 
DeDrift    & 132.6 & .03 & .19 & $132.8^{+}$ &
            .23 & .03 & .19 & .45     &
            – & – & –                     &
            19.17 & .03 & .55 & 19.75 \\ 
LIRE       & 44.2 & .03 & .38 & $44.61^{*+}$ &
            .15 & .05 & .11 & .31     &
            – & – & –                     &
            9.08 & .02 & .21 & $9.32^{*}$ \\ 
ScaNN      & 50.27 & 1.75 & 0 & $52.02^{+}$ &
            .41 & .21 & 0 & .62     &
            2.97 & 0 & 2.97             &
            6.70 & .09 & 0 & 6.79 \\ 
Faiss-HNSW & 14.65 & .18 & 0 & 14.83 &
            – & – & – & –                 &
            1.9 & 0 & 1.9             &
            1.27 & 1.38 & 0 & 2.64 \\ 
DiskANN    & 12.11 & .32 & 0 & 12.43 &
            .22 & 1.53 & 0 & $1.75^{*}$     &
            1.16 & 0 & 1.16             &
            .81 & .48 & 0 & 1.28 \\ 
SVS        & 20.54 & .57 & 0 & $21.11^{*}$ &
            .29 & 2.32 & 0 & 2.61     &
            .33 & 0 & \textbf{.33}             &
            2.11 & .24 & 0 & $2.35^{*}$ \\ \hline
\end{tabular}
\caption{\edit{Total workload time breakdown} \textbf{in hours}.  \textbf{S}: search, \textbf{U}: update, \textbf{M}: maintenance, \textbf{T}: overall total. Recall target = 90\% and k=100. Search queries are processed one-at-time, updates are processed in batches, maintenance is conducted after each batch of search or update operations. $^{*}$Denotes the method is unable to meet the recall target with static query parameters. $^{+}$Denotes the method did not finish in a 24 hour time budget, for these we estimate the runtime based on a $10\%$ subsample of search queries.}
\label{tab:workload_comparison}
\end{table*}

We configure the main parameters of Quake and the baselines as follows. We disable vector quantization/compression for all baselines, as not all baselines support it. For partitioned indexes we use $sqrt(|X_0|)$ partitions where $|X_0|$ is the initial number of vectors in the workload. For the graph indexes, we use a graph degree of 64. For LIRE and Quake, we set the partition refinement radius $r=50$. For Quake we use a single level of partition, set $\tau=250ns$, use one iteration of k-means for refinement and set $f_M$ between 1\%-10\%. All systems use 16 threads for updates and maintenance (if applicable). SCANN, DiskANN, and SVS perform maintenance eagerly during an update, therefore we do not measure maintenance time separately from update time. We consider maintenance after each operation for all methods. Throughout all experiments, indexes search parameters are tuned to achieve an average of 90\% recall for $k=100$ across the workloads. 

\subsection{End-to-End Evaluation}

\paragraph{Comparison with Baselines} Table~\ref{tab:workload_comparison} shows that Quake consistently achieves lower search, update and total time on all workloads. On the \textsc{Wikipedia-12M} workload, where the dataset grows over time and partitions can become unbalanced, the multi-threaded \edit{Quake-MT} takes 1.53 hours to process searches, \edit{while single-threaded Quake-ST takes 9.48 hours}. In contrast, Faiss-IVF climbs to 165 hours due to the lack of maintenance, DeDrift reaches 132 hours despite its rebalancing efforts, LIRE is unable to meet the recall target and takes 44 hours and SCANN performs similarly with poor update latency due to over-eager maintenance applied during updates. Even the best-performing graph-based method, DiskANN, takes 12 hours. Thus, \edit{Quake-MT} is $8\times$ faster to search than the strongest baseline on this workload derived from real-world access patterns.

On the \textsc{OpenImages-13M} workload, which includes both insertions and deletions, \edit{Quake's} multi-threaded and single-threaded search times are .03 and .14 hours respectively. The best competing approach, DiskANN, records .22 hours, making Quake-MT $7.3\times$ and Quake-ST $1.6\times$ faster. Faiss-HNSW does not support deletions so it is omitted. Both SVS's and DiskANN’s delete consolidation is expensive, leading to orders of magnitude higher update latency than partitioned indexes, illustrating that graph-based indexes struggle with dynamic operations. Quake’s continuous maintenance keep partitions balanced, achieving low latency and stable recall.

For the static, read-only, \textsc{MSTuring10M-RO} workload, Quake’s maintenance improves the index structure even without data changes, adapting partitions to the query pattern. For \edit{Quake-MT}, this yields a search time of .63 hours and \edit{Quake-ST} takes 2.43 hours to conduct the search. However, the \textsc{MSTuring10m} dataset is especially challenging for partitioned indexes, as they need to scan roughly $10\%$ of all partitions in order to meet the recall target. In contrast, the well-optimized SVS library exhibits a superior search time of .33 hours, demonstrating that in static settings, well-optimized graph indexes are strong competition.

On \textsc{MSTuring10M-IH}, where the dataset grows from one to ten million vectors, \edit{Quake-MT} achieves a total search time of .54 hours. DiskANN, the second-best performer, has a search time of .81 hours, making Quake-MT $1.5\times$ faster due to intra-query parallelism. However, single-threaded Quake is $2.6\times$ slower than DiskANN, further illustrating the search efficiency of graph indexes. The other baselines fail to maintain the recall target or suffer from high latency due to their static parameters and inability to prevent partition skew.

Overall, these results demonstrate that Quake’s combination of adaptive partition scanning, incremental maintenance, and NUMA-aware parallelism consistently delivers low-latency queries at the desired recall. Systems without maintenance (Faiss-IVF) suffer from skew-induced latency increases, those tied to static search parameters (LIRE) struggle to maintain recall without incurring higher query times, and graph-based methods (Faiss-HNSW, DiskANN) face substantial overheads when handling updates and deletions. By integrating these components, Quake matches the low update cost of partitioned indexes while outperforming graph indexes in search latency in dynamic workloads. Our design is an advancement to the state-of-the-art, providing stable, efficient performance across diverse, and evolving workloads.

\begin{figure}[h]
\centering
\includegraphics[width=0.43\textwidth]{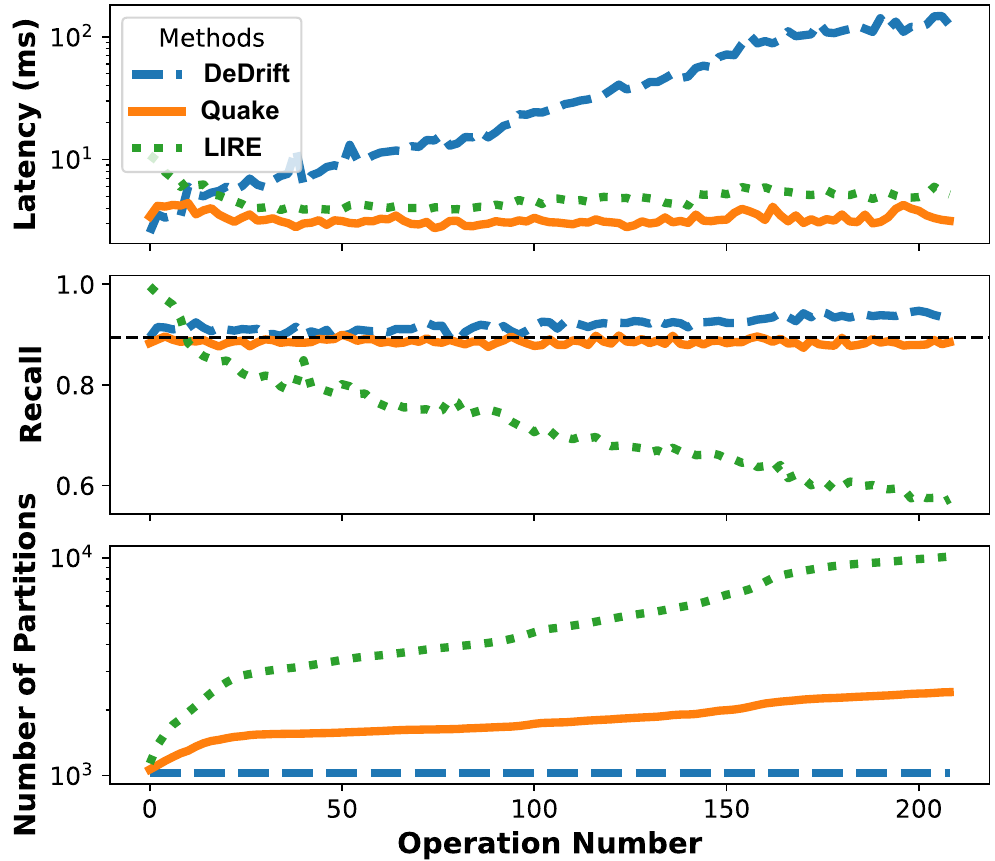}
\caption{Comparison of single-threaded search latency, recall and number of partitions for Quake vs. maintenance approaches LIRE and DeDrift on \textsc{Wikipedia-12M}. Quake maintains stable latency and recall throughout the workload.}
\label{fig:partitioned_maintenance}
\end{figure}

\paragraph{Comparison with Partitioned Index Maintenance Methods} 
Here we perform a detailed comparison with LIRE and DeDrift, measuring the latency, recall, and number of partitions over time on the \textsc{Wikipedia-12M} workload. For a fair comparison, we use a single-thread to highlight the advantages of APS and maintenance in Quake. The results are shown in Figure \ref{fig:partitioned_maintenance}. First looking at recall, we see that Quake maintains a stable recall of near 90\%, while LIRE's recall degrades over time as it uses a static nprobe. DeDrift's recall stays relatively constant, as it does not adjust the number of partitions and therefore does not need to adjust nprobe. However, when turning our attention to latency, we see that Quake has near-constant stable latency, even as the dataset grows, while DeDrift's latency increases significantly with time. In terms of the number of partitions, we see DeDrift stays constant while Quake and Lire increase by $2.5\times$ and $10\times$ respectively. LIRE uses significantly more partitions because it uses size thresholding to determine when to split, regardless of whether a given partition is hot or not. Quake on the other hand only splits partitions if their contribution to the cost model is high, allowing for more efficient maintenance. These results show that Quake's approach to maintenance is superior to existing methods for partitioned index maintenance in minimizing query latency and recall stability.

\begin{table}[h]
\footnotesize
\centering
\caption{Ablation Study on \textsc{Wikipedia-12M} showing mean search latency and the standard deviation of recall.}
\label{tab:ablation_study}
\begin{tabular}{|l|c|c|}
\hline
\textbf{Configuration} & \textbf{Search Latency} & \textbf{Recall Std.} \\ \hline
Quake-MT            & 0.53 ms      &  .008              \\ \hline
Quake-MT w/o APS            & 0.50 ms     &  .025                \\ \hline
Quake-ST           & 3.28 ms    &  .005                 \\ \hline
Quake-ST w/o APS       & 3.18 ms      &  .025             \\ \hline
Quake-ST w/o Maint/APS & 45.20 ms   &  .014                   \\ \hline
\end{tabular}
\end{table}

\paragraph{Wikipedia-12M Ablation}
To quantify the contributions of Quake components, we disabled key features and measured the impact on \textsc{Wikipedia-12M} workload in Table \ref{tab:ablation_study}.  We see that disabling APS has little impact on the query latency, as Quake can achieve a low latency even in the static nprobe setting. However, APS provides significantly more recall stability, as evidenced by the increase in standard deviation when APS is disabled. Disabling NUMA-aware multi-threading, however, shows a $6\times$ increase in query latency, demonstrating the benefit of parallelization of partition scans. Finally, we disable maintenance and see a significant increase in latency, similar to the latency of Faiss-IVF; here partitions become extremely imbalanced due to the skew in the workload (see Figure \ref{fig:wiki_combined}) causing queries to scan more vectors and therefore increasing latency. This further demonstrates the necessity for maintenance for dynamic workloads. In conclusion, each piece of Quake contributes to its performance in terms of both recall stability and minimal query latency.

\begin{figure}[h]
\centering
\includegraphics[width=0.4\textwidth]{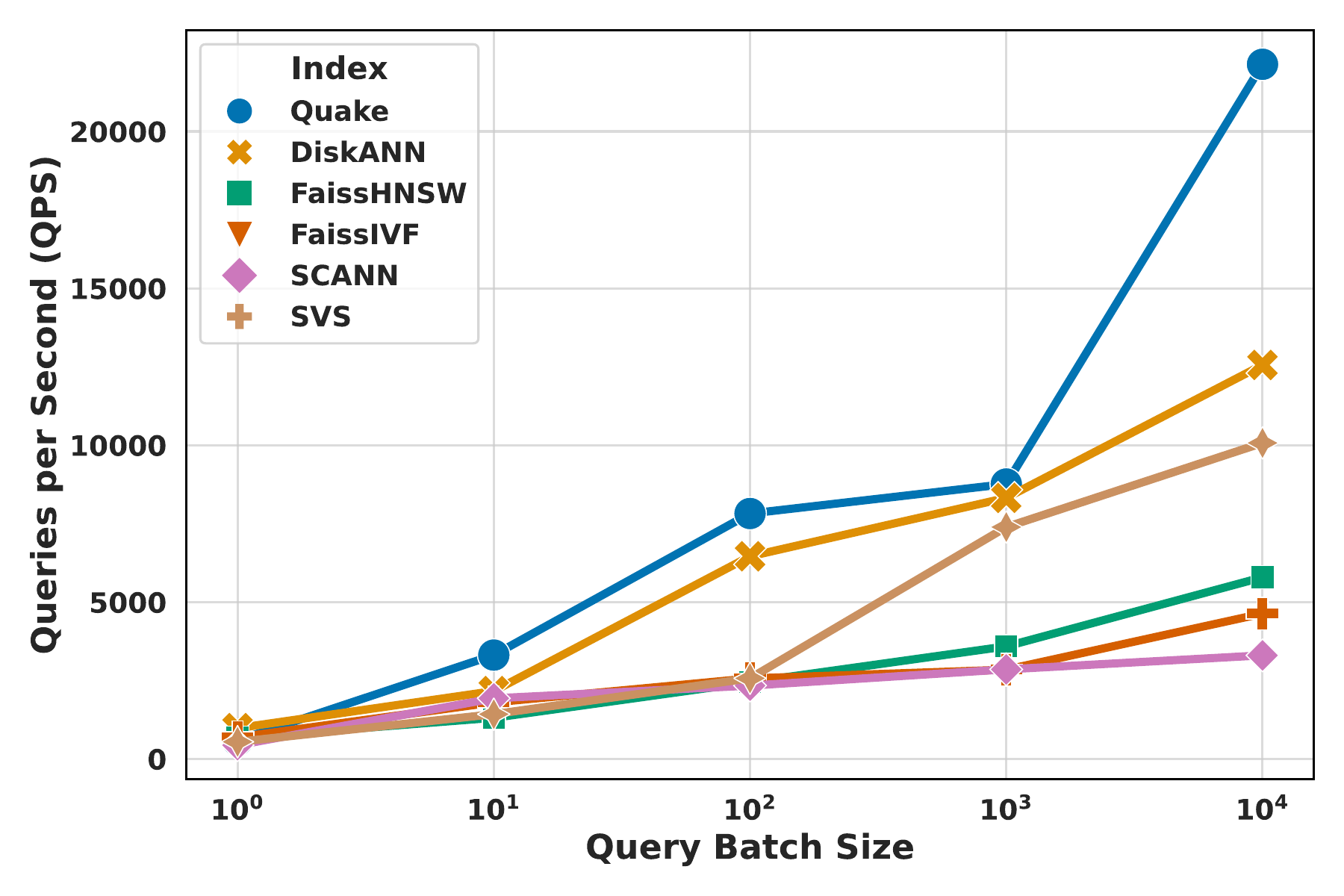}
\caption{Multi-query evaluation on \textsc{Wikipedia-12M} with 10,000 search queries. QPS @ recall=90\% is measured for all baselines while varying the batch size. All methods use 16 threads to process queries.}
\label{fig:qps_vs_batch_size}
\end{figure}

\edit{\subsection{Multi-Query Processing}}
Here we compare the search performance of Quake and baseline methods in a static batched query setting. Figure \ref{fig:qps_vs_batch_size} shows the QPS at a recall of 90\%, varying the number of queries in a batch on the \textsc{Wikipedia-12M} workload. The dataset includes all 12M vectors, with 10,000 queries sampled according to Wikipedia page views from December 2021. For Quake, FaissIVF, and SCANN, we use 3,500 partitions; for FaissHNSW, SVS, and DiskANN, we set the graph degree to 64. All experiments use 16 threads for query processing. Quake employs the multi-query execution policy in \cite{mohoney_high-throughput_2023} and \cite{pound2025micronn}, grouping queries by the partitions they access and scanning each partition exactly once per batch in parallel.

Quake consistently outperforms all baselines across every batch size, with an increasing advantage as batch sizes grow. At the largest batch size (10,000 queries), Quake achieves a $6.7\times$ speedup over FaissIVF and SCANN. This performance gain stems from Quake's efficient multi-query execution strategy, where it scans each partition once per batch, in contrast to FaissIVF and SCANN, which scan partitions individually per query. Compared to DiskANN, the strongest graph-based competitor, Quake still maintains a substantial $1.8\times$ speedup. These results demonstrate that Quake delivers high performance not only in single-query scenarios but also in large multi-query workloads.

%  showing that Quake exhibits superior performance even with limited resources.
% explosion
% exceptional 

\subsection{Scalability}

\begin{figure}[ht!]
    \centering
    \begin{subfigure}{0.23\textwidth}
        \centering
        \includegraphics[width=\textwidth]{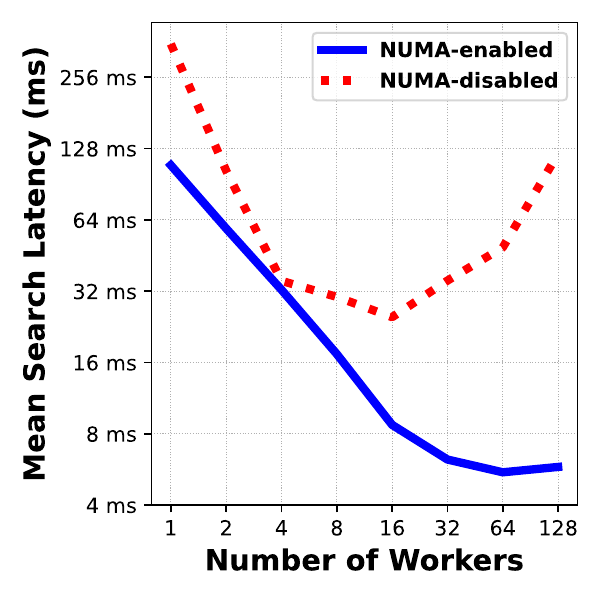}
        \caption{Mean Search Latency}
    \end{subfigure}
    \hfill
    \begin{subfigure}{0.23\textwidth}
        \centering
        \includegraphics[width=\textwidth]{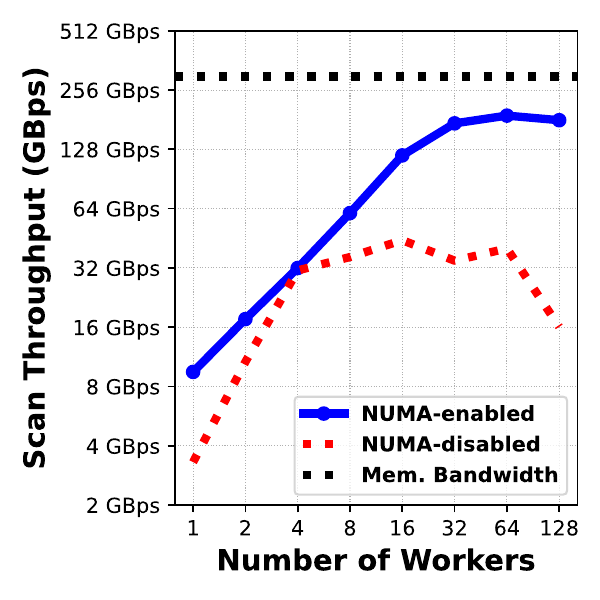}
        \caption{Scan Throughput}
    \end{subfigure}
    \caption{\textsc{MSTuring100M}: Scaling the number of threads with and without NUMA.}
    \label{fig:scalability_threads}
\end{figure}

We tested Quake’s parallel scalability by varying the number of threads. In Figure~\ref{fig:scalability_threads} we measure the mean search latency and scan throughput (bytes scanned / query latency) on \textsc{MSTuring100M} to reach a recall of 90\%. Note that this dataset has 100 million vectors and is $10\times$ larger than the datasets we compared against previously. We compare our NUMA-aware parallelism with one in which NUMA is disabled. For both configurations, we see near linear scalability up to around 8 workers, where the non-NUMA latency performs best (28ms). The NUMA configuration however further improves and at 64 workers achieves a latency of 6ms. Looking at the scan throughput, we see that NUMA achieves a peak throughput of 200GBps. We do not completely saturate memory bandwidth due to other overheads involved in query processing (topk sorting, memory allocations, coordination). In conclusion, NUMA-aware intra-query parallelism is an effective mechanism for decreasing query latency by utilizing the full memory capabilities of multi-core machines.

\begin{table}[h]
\footnotesize
\centering
\caption{\edit{Early-termination methods on \textsc{Sift1M} with a partitioned index with 1000 partitions. Each row shows the average recall, nprobe, and 
mean per-query latency in milliseconds over 10000 queries after tuning for a specific recall target for $k=100$. We also report the total offline tuning time in seconds, where APS needs \emph{no} offline
tuning.}}
\label{tab:early_term}
\begin{tabular}{lcccccc}
\toprule
Method & Target & Recall & $nprobe$ & Latency & Offline Tuning \\
\midrule
APS         & 80\% & 82.1\% & 11.8 & 0.34ms & \textbf{0} \\
            & 90\% & 91.2\% & 20.2 & 0.48ms & \textbf{0} \\
            & 99\% & 98.9\% & 50.1 & 0.96ms & \textbf{0} \\
\addlinespace
Auncel \cite{zhang2023fast}      & 80\% & 85.7\% & 16.4 & 0.41ms & 66.3s \\
            & 90\% & 98.1\% & 73.8 & 1.29ms & 73.8s \\
            & 99\% & 99.7\% & 95.9 & 1.61ms & 83.2s \\
\addlinespace
SPANN \cite{chen_spann_nodate}      & 80\% & 81.6\% & 11  & 0.31ms & 173s \\ 
            & 90\% & 90.2\% & 19  & 0.43ms & 183s \\
            & 99\% & 99.0\% & 70  & 1.07ms & 259s \\
\addlinespace
LAET \cite{li2020improving}       & 80\% & 81.3\% & 10.5 & 0.29ms & 81s \\
            & 90\% & 90.5\% & 18.2 & 0.42ms & 104s \\
            & 99\% & 99.0\% & 58.3 & 1.03ms & 232s \\
\addlinespace
Fixed & 80\% & 81.7\% & 11  & 0.33ms & 318s \\
            & 90\% & 90.3\% & 19  & 0.44ms & 330s \\
            & 99\% & 99.0\% & 65  & 1.16ms & 424s \\
\addlinespace
Oracle      & 80\% & 83.3\% & 11.5 & 0.29ms & 320s \\
            & 90\% & 92.4\% & 19.3 & 0.41ms & 331s \\
            & 99\% & 99.2\% & 42.0 & 0.74ms & 368s \\
\bottomrule
\end{tabular}
\end{table}

\edit{\subsection{Comparison with Early Termination Methods}}
Table \ref{tab:early_term} compares early-termination methods on \textsc{Sift1M}, highlighting the tradeoff between query latency, tuning time, and recall, we do not include ground truth generation time in the tuning time for the baselines. APS analytically estimates recall at query time, \textbf{eliminating offline tuning entirely}, while achieving latency within 30\% of the oracle across all recall targets. Fixed selects a static $nprobe$ per target via an expensive offline binary search (up to 424s), and SPANN similarly performs a binary search  and tunes a centroid-distance threshold; both closely match recall targets but incur higher latency at 90\% and 99\% recall. LAET trains a per-query prediction model, incurring moderate tuning overhead (81–232s), and matches recall targets with slightly higher latency compared to APS. Auncel is the most similar method to APS, as it aims to analytically estimate recall using partition intersection volumes, however its volume estimation requires calibration, and it is a conservative method, overshooting recall. We tune Auncel by binary searching a geometric parameter ($a$), overshooting recall significantly (up to 8.1 pp) and increasing latency by up to 169\% compared to APS. Finally, the Oracle, which scans the minimal amount of partitions per-query, serves as a practical lower bound on achievable latency, though with prohibitively high tuning cost. The tuning overhead of the baselines demonstrates a significant burden for online scenarios where the queries, data, and index change. The tuning burden worsens at scale, where ground truth generation cost grows linearly with the data size, and running queries multiple times to conduct binary search takes longer. APS thus provides near-optimal performance without tuning overhead, matching or exceeding all baselines.

\begin{table}[ht]
\footnotesize
\centering
\caption{
\textsc{Sift10M}: Recall and per‐level search latency ($\ell_{0}$, $\ell_{1}$, total) for a single‐level baseline (L$_0$: 40,000 partitions; L$_1$: 1 partition) and two‐level index (L$_0$: 40,000 partitions; L$_1$: 500 partitions), where recall targets are varied at each level. The single‐level baseline is the first row of each $\tau_{r}(0)$ block.
}
\label{tab:two_level_results}
\begin{tabular}{cccccc}
\toprule
$\tau_{r}(0)$ & $\tau_{r}(1)$ & Recall & $\ell_{0}$ (ms) & $\ell_{1}$ (ms) & Total (ms) \\
\midrule
\multirow{7}{*}{\centering 80\%} 
  & \textemdash & 81.2\% & 2.07 & 4.85 & 6.92 \\
\cmidrule(lr){2-6}
  & 80\%       & 74.8\% & 1.44 & 0.72       & 2.16 \\
  & 90\%       & 78.3\% & 1.56 & 1.19       & 2.75 \\
  & 95\%       & 80.1\% & 1.67 & 1.69       & 3.37 \\
  & 99\%       & 81.0\% & 1.75 & 2.57       & 4.33 \\
  & 100\%      & 81.1\% & 1.82 & 3.81       & 5.63 \\
\midrule
\multirow{7}{*}{\centering 90\%} 
  & \textemdash & 91.3\% & 2.85 & 4.89 & 7.86 \\
\cmidrule(lr){2-6}
  & 80\%       & 84.1\% & 2.07 & 0.77       & 2.84 \\
  & 90\%       & 88.2\% & 2.26 & 1.24       & 3.50 \\
  & 95\%       & 90.1\% & 2.38 & 1.72       & 4.10 \\
  & 99\%       & 91.0\% & 2.48 & 2.60       & 5.08 \\
  & 100\%      & 91.2\% & 2.62 & 3.88       & 6.50 \\
\midrule
\multirow{7}{*}{\centering 99\%} 
  & \textemdash & 99.0\% & 4.82 & 5.5 & 10.3 \\
\cmidrule(lr){2-6}
  & 80\%       & 91.4\% & 4.11 & 0.77       & 4.88 \\
  & 90\%       & 96.0\% & 4.59 & 1.27       & 5.86 \\
  & 95\%       & 97.7\% & 4.80 & 1.75       & 6.55 \\
  & 99\%       & 98.7\% & 5.08 & 2.65       & 7.74 \\
  & 100\%      & 98.9\% & 5.28 & 3.93       & 9.21 \\
\bottomrule
\end{tabular}
\end{table}

\subsection{Multi-Level Recall Estimation}
Here we evaluate the effectiveness of APS in a two-level partitioned index by measuring the impact of varying per-level recall targets on overall recall. Table~\ref{tab:two_level_results} reports results on \textsc{Sift10M} using 40{,}000 partitions at level $L_0$ and 500 partitions at level $L_1$. We use an initial search fraction of $f_M = 1.5\%$ at $L_0$ and 25\% at $L_1$.

We observe that overly aggressive early termination at $L_1$ by setting $\tau_r(1)$ too low leads to a clear degradation in total recall. For instance, at $\tau_r(0)=90\%$, reducing $\tau_r(1)$ from 99\% to 80\% lowers overall recall from 91.0\% to 84.1\%. This confirms that accurate recall estimation at the upper level is necessary to maintain accurate end-to-end recall estimates. These results justify our design decision in Section~\ref{sec:aps-algo} to fix $\tau_r(1)=99\%$ in multi-level configurations so that only the base recall target $\tau_r(0)$ needs to be provided.

In addition, the two-level index substantially reduces centroid scanning overhead. The single-level baseline must evaluate distances to all 40{,}000 centroids per query. In contrast, the two-level configuration performs an approximate search over the centroids. For example, at a recall target of $\tau_r(0) = 90\%$ with $\tau_r(1) = 99\%$ the total query latency drops from 7.86ms to 5.08ms, a 35\% reduction, driven mainly by the drop in $L_1$ latency from 4.89ms to 2.60ms.

\edit{\subsection{Maintenance Ablation}}
\label{sec:abl_maintenance}

To understand the effectiveness of the primary components of adaptive incremental maintenance (cost-model, partition refinement, and rejection), we replay a dynamic \textsc{Sift1M} trace (30\% inserts, 20\% deletes, 50\% queries) with different components disabled. All methods use a single-thread and search using APS with $k{=}100$ and a 90\% recall target. We also include LIRE as a baseline. Table \ref{tab:abl_sift} reports cumulative times in seconds. For configurations with refinement we use a refinement radius of $r_f=50$. The full Quake policy delivers the lowest search cost (86s) while meeting the recall target. If we keep the cost model but skip refinement (NoRef), maintenance time decreases significantly from 21s to 5s, yet recall slips by 2.4 pp and the search time increase by 15.4s. This shows that while refinement is the dominant cost in maintenance, it is necessary for minimizing search latency. Disabling the cost model and instead using size-based thresholding (NoCost) shows why naive size thresholds are inadequate: search time rises 8\% despite similar maintenance effort. The rejection mechanism is critical; once removed (NoRej), recall collapses to 66\% even though search and maintenance appear cheap. LIRE, which relies solely on size thresholding, is 17\% slower in search latency, confirming that the cost model, rejection mechanism, and partition refinement are essential for maintaining both index performance and quality.

\begin{table}[h]
\footnotesize
\centering
\caption{\edit{Maintenance ablation on the \textsc{Sift1M} workload. Times are cumulative (in seconds) over the course of the workload. Recall is averaged over all queries.}}
\label{tab:abl_sift}
\begin{tabular}{lrrrr}
\toprule
Maintenance Variant & Search & Update & Maint. & Recall \\
\midrule
Quake (Full)           & 86.3s & 21.7s & 21.4s & 90.5\% \\
NoRef            & 101.7s & 22.2s &  5.2s & 88.1\% \\
NoRef+NoRej      & 85.5s  & 21.0s &  1.0s & 73.0\% \\
NoRej            & 84.2s  & 19.6s & 18.5s & 66.2\% \\
NoCost           & 93.5s  & 20.0s & 20.4s & 90.1\% \\
NoCost+NoRef     & 100.7s & 21.2s &  0.8s & 87.9\% \\
\midrule
LIRE             & 100.5s & 21.2s & 11.9s & 90.0\% \\
\bottomrule
\label{tab:maintenance_ablation}
\end{tabular}
\end{table}

\section{\edit{Discussion}}

Here we offer a discussion of Quake’s system parameters, how the design extends to new hardware and use-cases.

\subsection{Setting System Parameters}
Quake exposes a few search and maintenance parameters, which we fix across all workloads unless otherwise stated. These defaults give stable performance with minimal tuning.

\paragraph{Search parameters.} The \textit{initial candidate fraction} $f_M$ determines the number of partitions to consider in APS. It has the largest impact on performance. If set too low, APS may not meet the recall target; if too high, the initial scan dominates latency. We set this between 1\% and 10\%. In future work, we aim to remove this parameter entirely. The \textit{number of worker threads} is best set to the number of physical cores. For large problem sizes, Quake scales linearly with thread count until memory bandwidth is saturated (Figure~\ref{fig:scalability_threads}). The \textit{recompute threshold} controls how often APS updates its recall estimate. We set this to 1\%, which avoids unnecessary recomputation with negligible impact on recall (Table~\ref{tab:recall_overhead}).

\paragraph{Maintenance parameters.} The \textit{split/merge threshold} $\tau$ sets the minimum predicted latency improvement required to trigger a split or delete. We set $\tau = 250$ns. Lower values increase maintenance cost and may cause over-splitting; higher values can allow imbalance to persist. The \textit{split access scaling parameter} $\alpha$ estimates the drop in access frequency after a split. We fix $\alpha = 0.9$, which worked well across all benchmarks. If maintenance tuning is needed, we recommend keeping $\alpha$ fixed and adjusting $\tau$. \textit{Refinement} is controlled by two parameters: the refinement radius $r_f$ (number of nearby partitions considered) and the number of refinement iterations. We use one iteration over the 50 nearest partitions. From our ablation study, disabling refinement reduces maintenance time by 75\% but increases query latency and reduces recall (Table~\ref{tab:maintenance_ablation}). The \textit{window size} for access frequency statistics is set equal to the maintenance interval. For example, if maintenance runs every 100{,}000 queries, the window also spans 100{,}000 queries. Smaller windows adapt faster but are more volatile.

Most parameters are fixed across workloads. In practice, only the initial candidate fraction $f_M$ and the maintenance threshold $\tau$ benefit from tuning; however, their defaults are sufficient for the workloads we evaluated.

\subsection{Deployment Considerations}
Quake’s design can be extended to support a range of hardware and use-cases through minor changes to the cost model and APS logic. We reserve these extensions for future work.

\paragraph{Heterogeneous Hardware} To adapt to devices with varying scan throughput such as CPUs, GPUs, or disk-backed storage, the cost model can be modified by profiling per-device scan latency (e.g., $\lambda_{\text{CPU}}(s)$, $\lambda_{\text{GPU}}(s)$, $\lambda_{\text{Disk}}(s)$) and updating partition cost estimates accordingly.

\paragraph{Distributed Environments} In a distributed setting, each machine can run APS and maintenance independently on its local partitions. The cost model can account for partition locality and inter-node variation. A separate load balancer would be required to assign and replicate partitions across machines in a way that minimizes total query cost.

\paragraph{Vector Compression} Vector compression techniques, such as Product Quantization~\cite{jegou_product_2011}, reduce scan cost. Quake can support compression by profiling scan latency over partitions of compressed vectors and updating $\lambda(s)$ in the cost model.

\paragraph{Filters} Filtered queries can be supported by scaling per-partition recall probabilities in APS, based on the estimated number of items that pass the filter in each partition. This will enable Quake to avoid scanning partitions unlikely to contain matching results while preserving recall targets.

\paragraph{Concurrency} The current implementation executes searches, updates, and maintenance serially. Quake can support concurrency through copy-on-write semantics, allowing background operations to build new index views while readers continue on the current one without blocking.

\section{Conclusion}
Experimental results show that Quake reduces query latency compared to baseline approaches under dynamic and skewed workloads, without requiring manual tuning. It achieves high recall, matching the performance of an oracle for setting the query parameter nprobe. Compared to existing partitioned indexes like Faiss and SCANN, Quake reduces query latency by A) adaptively maintaining index partitions and B) maximizing memory bandwidth during query processing. Compared to graph indexes like SVS, HNSW, and DiskANN, Quake offers more efficient indexing and updates while matching or reducing query latency. In summary, our evaluation shows Quake minimizes query latency while meeting recall targets on dynamic workloads with skewed access patterns.

\paragraph{Acknowledgments}
We would like to thank our shepherd, Nitin Agrawal, and the reviewers for their valuable feedback and efforts in making this a stronger paper. This work was supported by NSF grant CNS-2237306, Apple Scholars in AIML PhD Fellowship, and UW-Madison Hilldale Undergraduate Research Fellowship. This work was also supported by DARPA under the grant AIE DARPA-PA-22-01. The U.S. Government is authorized to reproduce and distribute reprints for Governmental purposes notwithstanding any copyright notation thereon. Any opinions, findings, and conclusions or recommendations expressed in this material are those of the authors and do not necessarily reflect the views, policies, or endorsements, either expressed or implied, of DARPA or the U.S. Government.

%-------------------------------------------------------------------------------
\bibliographystyle{plain}
\bibliography{sections/ref}

\begin{thebibliography}{10}

\bibitem{noauthor_qdrant_nodate}
Qdrant - {Vector} {Database}.
\newblock https://qdrant.tech/.

\bibitem{bigann}
Billion-scale approximate nearest neighbor search challenge: Neurips'21 competition track.
\newblock {https://big-ann-benchmarks.com/}, 2021.

\bibitem{pinecone-url}
Vector database for vector search | pinecone.
\newblock https://www.pinecone.io, 2024.
\newblock Accessed on December 4, 2023.

\bibitem{wiki}
Wikipedia:pageview statistics.
\newblock {https://en.wikipedia.org/wiki/Wikipedia:Pageview\_statistics}, 2024.

\bibitem{aguerrebere2024locally}
Cecilia Aguerrebere, Mark Hildebrand, Ishwar~Singh Bhati, Theodore Willke, and Mariano Tepper.
\newblock Locally-adaptive quantization for streaming vector search.
\newblock {\em arXiv preprint arXiv:2402.02044}, 2024.

\bibitem{baranchuk_dedrift_2023}
Dmitry Baranchuk, Matthijs Douze, Yash Upadhyay, and I.~Zeki Yalniz.
\newblock {DeDrift}: {Robust} {Similarity} {Search} under {Content} {Drift}, August 2023.
\newblock arXiv:2308.02752 [cs].

\bibitem{chen_spann_nodate}
Qi~Chen, Bing Zhao, Haidong Wang, Mingqin Li, Chuanjie Liu, Zengzhong Li, Mao Yang, and Jingdong Wang.
\newblock {SPANN}: {Highly}-efﬁcient {Billion}-scale {Approximate} {Nearest} {Neighbor} {Search}.

\bibitem{douze2024faisslibrary}
Matthijs Douze, Alexandr Guzhva, Chengqi Deng, Jeff Johnson, Gergely Szilvasy, Pierre-Emmanuel Mazaré, Maria Lomeli, Lucas Hosseini, and Hervé Jégou.
\newblock The faiss library, 2024.

\bibitem{grbovic2018real}
Mihajlo Grbovic and Haibin Cheng.
\newblock Real-time personalization using embeddings for search ranking at airbnb.
\newblock In {\em Proceedings of the 24th ACM SIGKDD International Conference on Knowledge Discovery \& Data Mining}, pages 311--320, 2018.

\bibitem{guo_accelerating_2020}
Ruiqi Guo, Philip Sun, Erik Lindgren, Quan Geng, David Simcha, Felix Chern, and Sanjiv Kumar.
\newblock Accelerating {Large}-{Scale} {Inference} with {Anisotropic} {Vector} {Quantization}.
\newblock In {\em Proceedings of the 37th {International} {Conference} on {Machine} {Learning}}, pages 3887--3896. PMLR, November 2020.
\newblock ISSN: 2640-3498.

\bibitem{haldar2019applying}
Malay Haldar, Mustafa Abdool, Prashant Ramanathan, Tao Xu, Shulin Yang, Huizhong Duan, Qing Zhang, Nick Barrow-Williams, Bradley~C Turnbull, Brendan~M Collins, et~al.
\newblock Applying deep learning to airbnb search.
\newblock In {\em Proceedings of the 25th ACM SIGKDD International Conference on Knowledge Discovery \& Data Mining}, pages 1927--1935, 2019.

\bibitem{hashemi2021neural}
Helia Hashemi, Aasish Pappu, Mi~Tian, Praveen Chandar, Mounia Lalmas, and Benjamin Carterette.
\newblock Neural instant search for music and podcast.
\newblock In {\em Proceedings of the 27th ACM SIGKDD Conference on Knowledge Discovery \& Data Mining}, pages 2984--2992, 2021.

\bibitem{jegou2010product}
Herve Jegou, Matthijs Douze, and Cordelia Schmid.
\newblock Product quantization for nearest neighbor search.
\newblock {\em IEEE transactions on pattern analysis and machine intelligence}, 33(1):117--128, 2010.

\bibitem{jegou_product_2011}
Herve Jégou, Matthijs Douze, and Cordelia Schmid.
\newblock Product {Quantization} for {Nearest} {Neighbor} {Search}.
\newblock {\em IEEE Transactions on Pattern Analysis and Machine Intelligence}, 33(1):117--128, January 2011.
\newblock Conference Name: IEEE Transactions on Pattern Analysis and Machine Intelligence.

\bibitem{Kuznetsova_2020}
Alina Kuznetsova, Hassan Rom, Neil Alldrin, Jasper Uijlings, Ivan Krasin, Jordi Pont-Tuset, Shahab Kamali, Stefan Popov, Matteo Malloci, Alexander Kolesnikov, Tom Duerig, and Vittorio Ferrari.
\newblock The open images dataset v4: Unified image classification, object detection, and visual relationship detection at scale.
\newblock {\em International Journal of Computer Vision}, 128(7):1956–1981, March 2020.

\bibitem{lee2014concise}
Yongjae Lee and Woo~Chang Kim.
\newblock Concise formulas for the surface area of the intersection of two hyperspherical caps.
\newblock {\em KAIST Technical Report}, 2014.

\bibitem{10.1145/2588555.2610507}
Viktor Leis, Peter Boncz, Alfons Kemper, and Thomas Neumann.
\newblock Morsel-driven parallelism: a numa-aware query evaluation framework for the many-core age.
\newblock In {\em Proceedings of the 2014 ACM SIGMOD International Conference on Management of Data}, SIGMOD '14, page 743–754, New York, NY, USA, 2014. Association for Computing Machinery.

\bibitem{li2020improving}
Conglong Li, Minjia Zhang, David~G Andersen, and Yuxiong He.
\newblock Improving approximate nearest neighbor search through learned adaptive early termination.
\newblock In {\em Proceedings of the 2020 ACM SIGMOD International Conference on Management of Data}, pages 2539--2554, 2020.

\bibitem{li2010concise}
Shengqiao Li.
\newblock Concise formulas for the area and volume of a hyperspherical cap.
\newblock {\em Asian Journal of Mathematics \& Statistics}, 4(1):66--70, 2010.

\bibitem{libtorch}
{LibTorch: PyTorch C++ API}.
\newblock {https://pytorch.org/cppdocs}.

\bibitem{liu2017related}
David~C Liu, Stephanie Rogers, Raymond Shiau, Dmitry Kislyuk, Kevin~C Ma, Zhigang Zhong, Jenny Liu, and Yushi Jing.
\newblock Related pins at pinterest: The evolution of a real-world recommender system.
\newblock In {\em Proceedings of the 26th international conference on world wide web companion}, pages 583--592, 2017.

\bibitem{liu2022monolith}
Zhuoran Liu, Leqi Zou, Xuan Zou, Caihua Wang, Biao Zhang, Da~Tang, Bolin Zhu, Yijie Zhu, Peng Wu, Ke~Wang, and Youlong Cheng.
\newblock Monolith: Real time recommendation system with collisionless embedding table.
\newblock In {\em 5th Workshop on Online Recommender Systems and User Modeling (ORSUM2022), in conjunction with the 16th ACM Conference on Recommender Systems}, 2022.

\bibitem{mageirakos2025cracking}
Vasilis Mageirakos, Bowen Wu, and Gustavo Alonso.
\newblock Cracking vector search indexes.
\newblock {\em arXiv preprint arXiv:2503.01823}, 2025.

\bibitem{10.1109/TPAMI.2018.2889473}
Yu~A. Malkov and D.~A. Yashunin.
\newblock Efficient and robust approximate nearest neighbor search using hierarchical navigable small world graphs.
\newblock {\em IEEE Trans. Pattern Anal. Mach. Intell.}, 42(4):824–836, April 2020.

\bibitem{mohoney2024incremental}
Jason Mohoney, Anil Pacaci, Shihabur~Rahman Chowdhury, Umar~Farooq Minhas, Jeffery Pound, Cedric Renggli, Nima Reyhani, Ihab~F Ilyas, Theodoros Rekatsinas, and Shivaram Venkataraman.
\newblock Incremental ivf index maintenance for streaming vector search.
\newblock {\em arXiv preprint arXiv:2411.00970}, 2024.

\bibitem{mohoney_high-throughput_2023}
Jason Mohoney, Anil Pacaci, Shihabur~Rahman Chowdhury, Ali Mousavi, Ihab~F. Ilyas, Umar~Farooq Minhas, Jeffrey Pound, and Theodoros Rekatsinas.
\newblock High-{Throughput} {Vector} {Similarity} {Search} in {Knowledge} {Graphs}.
\newblock {\em Proceedings of the ACM on Management of Data}, 1(2):1--25, June 2023.

\bibitem{mohoney2025quakeadaptiveindexingvector}
Jason Mohoney, Devesh Sarda, Mengze Tang, Shihabur~Rahman Chowdhury, Anil Pacaci, Ihab~F. Ilyas, Theodoros Rekatsinas, and Shivaram Venkataraman.
\newblock Quake: Adaptive indexing for vector search (technical report).
\newblock {\em arXiv preprint arXiv:2506.03437}, 2025.

\bibitem{mohoney2021marius}
Jason Mohoney, Roger Waleffe, Henry Xu, Theodoros Rekatsinas, and Shivaram Venkataraman.
\newblock Marius: Learning massive graph embeddings on a single machine.
\newblock In {\em 15th $\{$USENIX$\}$ Symposium on Operating Systems Design and Implementation ($\{$OSDI$\}$ 21)}, pages 533--549, 2021.

\bibitem{moody}
{moodycamel::ConcurrentQueue}.
\newblock {https://github.com/cameron314/concurrentqueue}.

\bibitem{simsimd}
{SimSIMD}.
\newblock {https://github.com/ashvardanian/SimSIMD}.

\bibitem{ni_diskann_2023}
Jiongkang Ni, Xiaoliang Xu, Yuxiang Wang, Can Li, Jiajie Yao, Shihai Xiao, and Xuecang Zhang.
\newblock {DiskANN}++: {Efficient} {Page}-based {Search} over {Isomorphic} {Mapped} {Graph} {Index} using {Query}-sensitivity {Entry} {Vertex}, November 2023.
\newblock arXiv:2310.00402 [cs].

\bibitem{okura2017embedding}
Shumpei Okura, Yukihiro Tagami, Shingo Ono, and Akira Tajima.
\newblock Embedding-based news recommendation for millions of users.
\newblock In {\em Proceedings of the 23rd ACM SIGKDD international conference on knowledge discovery and data mining}, pages 1933--1942, 2017.

\bibitem{pal2020pinnersage}
Aditya Pal, Chantat Eksombatchai, Yitong Zhou, Bo~Zhao, Charles Rosenberg, and Jure Leskovec.
\newblock Pinnersage: Multi-modal user embedding framework for recommendations at pinterest.
\newblock In {\em Proceedings of the 26th ACM SIGKDD International Conference on Knowledge Discovery \& Data Mining}, pages 2311--2320, 2020.

\bibitem{pound2025micronn}
Jeffrey Pound, Floris Chabert, Arjun Bhushan, Ankur Goswami, Anil Pacaci, and Shihabur~Rahman Chowdhury.
\newblock Micronn: An on-device disk-resident updatable vector database.
\newblock {\em arXiv preprint arXiv:2504.05573}, 2025.

\bibitem{10.14778/3015274.3015275}
Iraklis Psaroudakis, Tobias Scheuer, Norman May, Abdelkader Sellami, and Anastasia Ailamaki.
\newblock Adaptive numa-aware data placement and task scheduling for analytical workloads in main-memory column-stores.
\newblock {\em Proc. VLDB Endow.}, 10(2):37–48, October 2016.

\bibitem{qin2021mixer}
An~Qin, Mengbai Xiao, Yongwei Wu, Xinjie Huang, and Xiaodong Zhang.
\newblock Mixer: efficiently understanding and retrieving visual content at web-scale.
\newblock {\em Proceedings of the VLDB Endowment}, 14(12):2906--2917, 2021.

\bibitem{radford2021learningtransferablevisualmodels}
Alec Radford, Jong~Wook Kim, Chris Hallacy, Aditya Ramesh, Gabriel Goh, Sandhini Agarwal, Girish Sastry, Amanda Askell, Pamela Mishkin, Jack Clark, Gretchen Krueger, and Ilya Sutskever.
\newblock Learning transferable visual models from natural language supervision, 2021.

\bibitem{singh2021freshdiskann}
Aditi Singh, Suhas~Jayaram Subramanya, Ravishankar Krishnaswamy, and Harsha~Vardhan Simhadri.
\newblock Freshdiskann: A fast and accurate graph-based ann index for streaming similarity search.
\newblock {\em arXiv preprint arXiv:2105.09613}, 2021.

\bibitem{10.5555/3454287.3455520}
Suhas~Jayaram Subramanya, Devvrit, Rohan Kadekodi, Ravishankar Krishaswamy, and Harsha~Vardhan Simhadri.
\newblock {\em DiskANN: fast accurate billion-point nearest neighbor search on a single node}.
\newblock Curran Associates Inc., Red Hook, NY, USA, 2019.

\bibitem{soar_2023}
Philip Sun, David Simcha, Dave Dopson, Ruiqi Guo, and Sanjiv Kumar.
\newblock Soar: Improved indexing for approximate nearest neighbor search.
\newblock In {\em Neural Information Processing Systems}, 2023.

\bibitem{waleffe2023mariusgnn}
Roger Waleffe, Jason Mohoney, Theodoros Rekatsinas, and Shivaram Venkataraman.
\newblock Mariusgnn: Resource-efficient out-of-core training of graph neural networks.
\newblock In {\em ACM SIGOPS European Conference on Computer Systems (EuroSys)}, 2023.

\bibitem{wang2021milvus}
Jianguo Wang, Xiaomeng Yi, Rentong Guo, Hai Jin, Peng Xu, Shengjun Li, Xiangyu Wang, Xiangzhou Guo, Chengming Li, Xiaohai Xu, et~al.
\newblock Milvus: A purpose-built vector data management system.
\newblock In {\em Proceedings of the 2021 International Conference on Management of Data}, pages 2614--2627, 2021.

\bibitem{wang2018billion}
Jizhe Wang, Pipei Huang, Huan Zhao, Zhibo Zhang, Binqiang Zhao, and Dik~Lun Lee.
\newblock Billion-scale commodity embedding for e-commerce recommendation in alibaba.
\newblock In {\em Proceedings of the 24th ACM SIGKDD International Conference on Knowledge Discovery \& Data Mining}, pages 839--848, 2018.

\bibitem{wei2020analyticdb}
Chuangxian Wei, Bin Wu, Sheng Wang, Renjie Lou, Chaoqun Zhan, Feifei Li, and Yuanzhe Cai.
\newblock Analyticdb-v: A hybrid analytical engine towards query fusion for structured and unstructured data.
\newblock {\em Proceedings of the VLDB Endowment}, 13(12):3152--3165, 2020.

\bibitem{xu_spfresh_2023}
Yuming Xu, Hengyu Liang, Jin Li, Shuotao Xu, Qi~Chen, Qianxi Zhang, Cheng Li, Ziyue Yang, Fan Yang, Yuqing Yang, Peng Cheng, and Mao Yang.
\newblock {SPFresh}: {Incremental} {In}-{Place} {Update} for {Billion}-{Scale} {Vector} {Search}.
\newblock In {\em Proceedings of the 29th {Symposium} on {Operating} {Systems} {Principles}}, {SOSP} '23, pages 545--561, New York, NY, USA, October 2023. Association for Computing Machinery.

\bibitem{distmult}
Bishan Yang, Wen-tau Yih, Xiaodong He, Jianfeng Gao, and Li~Deng.
\newblock Embedding entities and relations for learning and inference in knowledge bases.
\newblock {\em arXiv preprint arXiv:1412.6575}, 2014.

\bibitem{zhang2023vbase}
Qianxi Zhang, Shuotao Xu, Qi~Chen, Guoxin Sui, Jiadong Xie, Zhizhen Cai, Yaoqi Chen, Yinxuan He, Yuqing Yang, Fan Yang, et~al.
\newblock $\{$VBASE$\}$: Unifying online vector similarity search and relational queries via relaxed monotonicity.
\newblock In {\em 17th USENIX Symposium on Operating Systems Design and Implementation (OSDI 23)}, pages 377--395, 2023.

\bibitem{zhang2023fast}
Zili Zhang, Chao Jin, Linpeng Tang, Xuanzhe Liu, and Xin Jin.
\newblock Fast, approximate vector queries on very large unstructured datasets.
\newblock In {\em 20th USENIX Symposium on Networked Systems Design and Implementation (NSDI 23)}, pages 995--1011, 2023.

\end{thebibliography}

\appendix
\section{Artifact Appendix}

\subsection*{Abstract}
This artifact provides the experimental setup for Quake to reproduce key results. See the artifact's README for full details.

\subsection*{Scope}
This artifact enables validation of Quake's core experimental findings, such as those related to Adaptive Partition Scanning (APS), NUMA-aware searching, and maintenance policies. Refer to the artifact's README for specific experiments.

\subsection*{Contents}
The artifact includes Python scripts (e.g., \texttt{experiment\_runner.py}, individual experiment \texttt{run.py} files), system installation scripts (\texttt{install.sh}), experiment configurations (\texttt{configs/}), Conda environment files, and the paper PDF. The README details the full directory structure and contents.

\subsection*{Hosting}
The artifact is part of the Quake repository in the \texttt{osdi2025} branch. located at \url{https://github.com/marius-team/quake/tree/osdi2025/test/experiments/osdi2025}. Use the latest commit on this branch.

\subsection*{Requirements}
Python 3.9+ and Conda are required. Tested on Linux (Ubuntu 22.04). Some experiments (e.g., NUMA evaluations) require a machine with NUMA for meaningful reproduction. The \texttt{install.sh} script lists system-level dependencies. Refer to the README for comprehensive requirements.

\subsection{Installation}
Follow the detailed installation instructions in the artifact's README. Options include a comprehensive setup using \texttt{install.sh} or a Quake-only Conda environment setup.

\subsection{Experiment Workflow}
All experiments are launched via \texttt{experiment\_runner.py} from the repository root. Experiments will typically download and prepare required datasets if not found locally. Experiments are run as follows: \\
\texttt{python3 -m test.experiments.osdi2025.experiment\_runner --experiment kick\_the\_tires -- config sift1m} \\
After execution, experiments print status updates and save results (e.g., CSV files, plots) to an output directory, as indicated in the console output. The artifact's README provides the complete command structure, detailed explanations of all parameters (including \texttt{--output-dir}), further examples, and a full summary of available experiments with their specific configurations.

%%%%%%%%%%%%%%%%%%%%%%%%%%%%%%%%%%%%%%%%%%%%%%%%%%%%%%%%%%%%%%%%%%%%%%%%%%%%%%%%
\end{document}